\documentclass{aa}                             
\usepackage{graphicx}
\usepackage{array}
\usepackage{lscape}                                
\usepackage{txfonts}
\usepackage{natbib}
\usepackage{longtable}
\usepackage{color}
\newcommand{\kms} {km\,s$^{-1}$}
\newcommand{\vsini} {$v$\,sin\,$i$}
\newcommand{\macro} {{\em macroturbulence}}
\newcommand{\macrot} {{\em macroturbulent}}
\newcommand{\vmicro} {$\xi_{\rm t}$}
\newcommand{\vmacro} {$v_{\rm m}$}
\newcommand{\TRT} {$\Theta_{\rm RT}$}
\newcommand{\TG} {$\Theta_{\rm G}$}
\begin{document}
%
\title{The IACOB project\thanks{Based on observations made with the Nordic 
       Optical Telescope, operated on the island of La Palma jointly by Denmark, 
       Finland, Iceland, Norway, and Sweden, in the Spanish Observatorio del 
       Roque de los Muchachos of the Instituto de Astrofisica de Canarias.}}
\subtitle{I. Rotational velocities in Northern Galactic O and early B-type stars revisited. 
The impact of other sources of line-broadening} 

\author{
S. Sim\'on-D\'{\i}az\inst{1,2} \& A. Herrero\inst{1,2}
}

\institute{
Instituto de Astrof\'isica de Canarias, E-38200 La Laguna, Tenerife, Spain.              
\and
Departamento de Astrof\'isica, Universidad de La Laguna, E-38205 La Laguna, Tenerife, Spain.
}
		   
\offprints{ssimon@iac.es}

\date{Submitted/Accepted}

\titlerunning{Rotational velocities in Northern Galactic OB-type stars}
\authorrunning{Sim\'on-D\'iaz \& Herrero}

%
\abstract{Stellar rotation is an important parameter in the evolution of
massive stars. Accurate and reliable measurements of projected rotational velocities 
in large samples of OB stars are crucial to confront the predictions of stellar 
evolutionary models with observational constraints.}
{We reassess previous determinations of projected rotational velocities (\vsini) in 
Galactic OB stars using a large, high quality spectroscopic dataset, and a strategy
which account for other sources of broadening appart from rotation affecting the 
diagnostic lines}
{We present a versatile and user friendly IDL tool --- based on a combined Fourier 
Transform (FT) + goodness of fit (GOF) methodology --- for the line-broadening
characterization in OB-type stars. We use this tool to (a) investigate the impact of 
\macrot\ and microturbulent broadenings on \vsini\ measurements, and (b) determine 
\vsini\ in a sample of $\sim$\,200 Galactic OB-type stars, also characterizing the 
amount of \macrot\ broadening (\vmacro) affecting the line profiles.}
{We present observational evidence illustrating the strengths and limitations of 
the proposed FT+GOF methodology for the case of OB stars. 
We confirm previous statements (based on indirect arguments or smaller samples) that 
the \macrot\ broadening is ubiquitous in the massive star domain. We compare the
newly derived \vsini\ in the case of O stars and early-B Supergiants and Giants
(where the effect of \macro\ is found to be larger) with previous determinations not accounting for this extra 
line-broadening contribution, and show that those cases with \vsini\,$\le$\,120~\kms 
need to be systematically revised downwards by $\sim$\,25 ($\pm$\,20)~\kms. We suggest 
that microturbulence may impose an upper limit below which \vsini\ and \vmacro\ could be incorrectly 
derived by means of the proposed methodology as presently used, and discuss the 
implications of this statement on the study of relatively narrow line massive stars.}
{An investigation of impact of the revised \vsini\ distributions on the predictions by
massive star evolutionary models is now warranted. Also, the reliability of \vsini\ measurements 
in the low \vsini\ regime, using a more precise description of the intrinsic profiles
used for the line-broadening analysis, needs to be further investigated.
}
%
\maketitle
%
%
\section{Introduction}\label{Sect1}
The stellar mass is the primary parameter determining the structure and 
evolution of a star and, traditionally, the chemical composition (viz. metallicity)
has been considered as the second most relevant parameter. However, it is already 15 
years since \cite{Mae00} highlighted that {\em rotation is indispensable for 
a proper modeling of the evolution for the upper Main Sequence stars} in their 
review about the evolution of rotating massive stars \citep[see also the more
recent review by][and references in both documents]{Lan12}.
Thanks to these and subsequent related studies we know that rotation in massive 
stars may play a role comparable to that of mass and metallicity in our interpretation
of the position occupied by these stars in the Hertzsprung-Russell Diagram.
Also, the inclusion of the effects of rotation (including rotational mixing) in 
massive star models has been crucial for the investigation of (a) the surface 
abundance pattern observed in massive stars \cite[e.g][]{Hun09, Prz10, Bou13}, 
and (b) the effect of very extreme
rotation in their evolution\footnote{Extreme rotation is predicted to be responsible
for homogeneous evolution \citep{Mae87}, in which the inner distribution of angular momentum
is importantly affected from the very early evolutionary stages.} and final fate 
\citep[long-duration gamma-ray burst and hypernovae, peculiar bright 
Type II and Type Ib/c supernovae, e.g.,][]{Geo09, Lan12}.

Even more recent works show the importance of binary interactions for the evolution
of massive stars and for the interpretation of population observations and synthesis 
population calculations \citep[see][and references therein]{San12, dMi13}.
These studies reveal that the present stellar rotation rate is a key parameter indicative of the 
past history of the binary systems. 

All the predictions by rotating stellar evolution models must be always supported
by observational constraints. Ideally, we would like to know the initial 
distribution of rotational velocities and their temporal evolution as a function of the 
other stellar properties (mass, luminosity, metallicity, multiplicity, etc.). However,
what we can actually measure is the projected rotational velocity (\vsini) of a certain star at given 
instant in its life. Therefore, to compensate this observational deficiency, \vsini\ 
measurements of large samples of O and early-B main sequence stars, as well as B Supergiants 
(B~Sgs) are of ultimate importance for our understanding of the massive stars and the 
stellar populations including (and often being dominated by) them.

One of the most straightforward and cheapest (from an observational point of view) 
ways to obtain information about (projected) rotational velocities in stars 
is based on the effect that rotation produces on the spectral line profiles 
(i.e. line-broadening). Following this approach, the analysis of large samples of 
Galactic O and early-B type stars via linewidth measurements \citep[e.g.,][]{Sle56, Con77} 
or cross-correlation techniques \citep[e.g.,][]{Pen96, How97}
provided a general overview of the rotational properties 
of stars in the upper part of the Hertzprung-Russell diagram. 

The \vsini\ measurements in all these works are, however, obtained under the assumption 
that rotation is the sole source of broadening. While the quantification of the 
global line-broadening is a relatively easy task, the translation of these quantities 
to actual projected rotational velocities in the case of O and B-type 
stars may become complex in certain situations \citep[see e.g., some notes in][and 
references therein]{How04}. The determination of actual \vsini\ becomes even more 
complicated with the increasing evidence that rotation is not the only broadening 
mechanism shaping the line-profiles of these stars. Already \cite{Str52} provided
several convincing arguments against a strict rotational interpretation of line-broadening
in this class of stars. \cite{Sle56}, \cite{Con77}, \cite{Pen96}, \cite{How97} also 
supported this statement based on the lack of O-type stars 
and early-B~Sgs with sharp absorption lines among their large analyzed samples.
More recently, the advent of high-quality spectroscopic observations (in terms 
of resolving power and signal-to-noise ratio) and its analysis by means of adequate 
techniques \citep[e.g.,][]{Rya02, Sim07} has allowed us to 
confirm the presence of an additional broadening mechanism shaping the line-profiles 
of these type of stars. It was called \macrot\ broadening at some point; 
however, it is easy to discard that this extra-broadening is produced by any type
of large scale turbulent motion \citep{Sim10}.
Nevertheless, for the sake of simplicity we will keep the name \macro\ in this paper.

Once the suspicions by Struve were confirmed, several questions needed to be 
treated: (a) is it possible to disentangle both broadening contributions from the 
line profiles?, (b) how are previous determinations of projected rotational velocities 
affected? and, last but not least (c) what is the physical origin of the anomalous 
non-rotational broadening?.

The last decade has witnessed good progress in the investigation of all these 
questions, specially in the case of B~Sgs \citep[e.g.,][]{Duf06, Lef07, Mar08, Aer09,
Fra10, Sim10}. However, a complete understanding of the impact of \macrot\ broadening 
in the whole massive star domain requires a 
comprehensive, homogeneous analysis of a large sample of high quality spectra of 
OB stars, also including early-B dwarfs and giants and, of course, O-type stars
of all luminosity classes. This was one of the initial drivers of the IACOB 
project (P.I. Sim\'on-D\'iaz), which aims at progressing in our knowledge of 
Galactic massive stars using a large, homogeneous, high-quality spectroscopic 
dataset and modern tools for the quantitative spectroscopic analysis of O and 
B-type stars.

Some preliminary results concerning 
the study of the projected rotational velocities and the \macrot\ broadening 
in the IACOB sample can be found in \cite{Sim12}, and references therein. 
Here, we present for the first time the complete study, also including a detailed 
description of the strengths and limitations of the methodology we have applied to 
disentagle both broadening contributions in a sample of $\sim$200 Northern Galactic 
O and early-B stars (covering all luminosity classes). We also refer the reader to
the recent study by \cite{Mar13}. In a parallel work to this one, these
authors provide observational constraints on the projected rotational velocities and
the amount of \macrot\ broadening as a function of fundamental parameters 
and stellar evolution from the analysis of a sample of 31 Southern Galactic O stars
(own new data) plus 86 OB Supergiants from the literature. 

The paper is structured as follows. In Sect. \ref{Sect2} a brief description of the
spectroscopic dataset used for this study is presented. The strategy we have followed
for the line-broadening characterization of our sample of O and B stars is described
in Sect. \ref{Sect3}. Special emphasis is made in the description
of the software we have used, the {\tt iacob-broad} tool, a versatile procedure 
developed and implemented by us in IDL,
that provides a complete set of quantities and visual information resulting from a 
combined Fourier transform plus a goodness of fit analysis of the line-broadening 
in OB stars. The analysis of the whole sample and the discussion of results can be 
found in Sect. \ref{Sect4}, while the main conclusions are summarized in Sect. \ref{Sect5}.

\section{Observations}\label{Sect2}

The spectroscopic observations considered for this study are part of the {\em IACOB 
spectroscopic database of Northern Galactic OB stars} \citep[last described in][]{Sim11a, Sim11b}.
This unique high-quality spectroscopic database has been compiled in the framework of 
the IACOB project. To date, the IACOB database comprises 1250
spectra of 153 and 97 Galactic O and early B-type stars, respectively, observable
from the Roque de los Muchachos observatory in La Palma (Spain). The spectra, having
a resolving power of 46000 and 23000, and a typical signal-to-noise (S/N) ratio above 150,
were compiled between November 2008 and January 2013 with the high-resolution 
FIbre-fed Echelle Spectrograph (FIES) attached to the Nordic Optical Telescope (NOT). 
The IACOB database has a multi-epoch character to investigate the binary/multiple 
nature of considered stars and the temporal variations in individual objects, with at 
least 3 spectra per observed target. In this 
study, we have only used a subsample of the spectra, discarding all those stars 
in which we have found signatures of multiplicity contaminating the spectrum (therefore considering only
apparently single and SB1 stars), and only considering 
the spectrum with the highest S/N ratio per star. 

All the spectra were homogeneously reduced using the FIES-tool pipeline, revised and normalized
using own procedures developed in IDL. The spectroscopic observation covers the full
wavelength range between 3700 and 7000 \AA\ without gaps; however, for this study we
have only used the spectral windows including the \ion{Si}{iii}\,$\lambda$4552
and \ion{O}{iii}\,$\lambda$5592 lines for the case of early-B and O-type stars, 
respectively.

\section{Methods}\label{Sect3}

Motivated by the investigation performed in this paper we have developed a versatile
and user friendly IDL tool for the line-broadening characterization of OB-type 
spectra: the {\tt iacob-broad}. This procedure allows us to determine \vsini\ and 
the \macrot\ broadening (\vmacro) under a variety of situations, and it is based on 
a combined Fourier Transform (FT) + goodness of fit (GOF) methodology.

These two methods are (independently) able to separate the rotational broadening contribution 
to the line profile from other line-broadening components, but they also present some 
caveats that must be taken into account for a correct interpretation of the results. We 
have found that a combined use of both techniques helps us to not only provide more reliable final estimates 
of \vsini\ and \vmacro\ (plus their associated uncertainties) but also to better understand the problematic 
cases and identify spurious solutions.

While the {\tt iacob-broad} tool has been used in several publications in the
last few years \citep[e.g.,][]{Sim10, 
DeG10, Aer13, Ram13, Sun13, Mar13}, it has not been formally 
described yet. In this section, we briefly describe the basics of the tool and explore the 
strengths and limitations of the proposed methodology. To this aim, we present a couple of formal tests carried 
out to investigate the effect that the assumed \macrot\ profile has on the GOF-based \vsini\ and 
\vmacro\ estimations (Sect.~\ref{Sect33}), and the possible consequences of neglecting
the effect of microturbulence on the determination of these two quantities (Sect.~\ref{Sect34}).

\subsection{The FT and GOF methods in a nutshell}\label{Sect31}

The FT method was first introduced by \cite{Car33} and, some years later, revisited and 
extended by \cite{Gra76,Gra05}. This technique is based (in its more simple form) on the 
identification of the first zero in the Fourier transform of the line-profile, and the 
direct translation of the associated frequency ($\sigma_{1}$) to the corresponding projected 
rotational velocity via the formula:
\begin{equation}\label{form1}
\frac{\lambda}{c}\,v\,$sin$\,i\, \sigma_{1} = 0.660.
\end{equation}
where $\lambda$ is the central line wavelength and c is the velocity of light.
This technique has been commonly used in the case of cool stars since the pioneering works 
by e.g. \cite{Wil69}, \cite{Gra73}, or \cite{Smi76}, but only marginally applied to 
the case of early OB-type stars before the first decade of the 
XXI$^{\rm st}$ century. After the work by \cite{Ebb79}, it was \cite{Sim07} who first applied 
this technique in a systematic way to the OB star domain. Since then, the FT method has been 
increasingly used for the study of this type of stars \citep[e.g.,][]{Duf06, 
Lef07, Hun08, Mar08, Fra10, Lef10, Bou12, Gru12, Duf13, Ram13, Mar13}. 

Being a powerful technique, the FT method must be used with special care when the rotational 
broadening is comparable to (or smaller than) the resolving power of the spectrum and/or the 
intrinsic\footnote{In this context, the intrinsic line profile includes the natural, thermal,
and Stark broadening.} broadening of the line. In addition, three other caveats must be kept in mind when 
extracting information about \vsini\ from the FT of the line-profile: (a) when the so-called 
\macrot\ broadening is comparable or larger than the 
rotational broadening and the S/N ratio of the line is not large enough, the position of the first 
zero could be hidden below the white-noise in the Fourier space, resulting in an apparent 
displacement to smaller frequencies (i.e. larger \vsini). An example of this 
effect can be found in \cite{Sim07}. Moreover, even when the line quality
is optimal, we should keep in mind the word of caution indicated by \cite{Gra05}: when 
macroturbulence dominates over rotation the sidelobe structure of the Fourier transform may 
be affected; (b) if the line is asymmetric and/or the final line-profile is not 
the result of a convolution of the various broadening agents, the first zero of the FT could not 
be related to the actual projected rotational velocity. An example of this caveat is found in
the work by \cite{Aer09}. If the \macrot\ broadening has a pulsational origin (a possibility
investigated by these authors in B~Sgs) then 
\vsini\ determinations could be seriously underestimated by using 
a simple parameter description for \macro\ rather than an appropriate pulsational model 
description to fit the line profiles; and c) as indicated by \cite{Gra73}, microturbulent 
broadening also adds zeroes to the FT at frequencies associated to low \vsini\ values which could 
be misinterpreted as zeroes produced by rotation (see also discussion in Sect.~\ref{Sect34}).

\begin{figure}[t!]
\centering
\includegraphics[width=8.5 cm,angle=0]{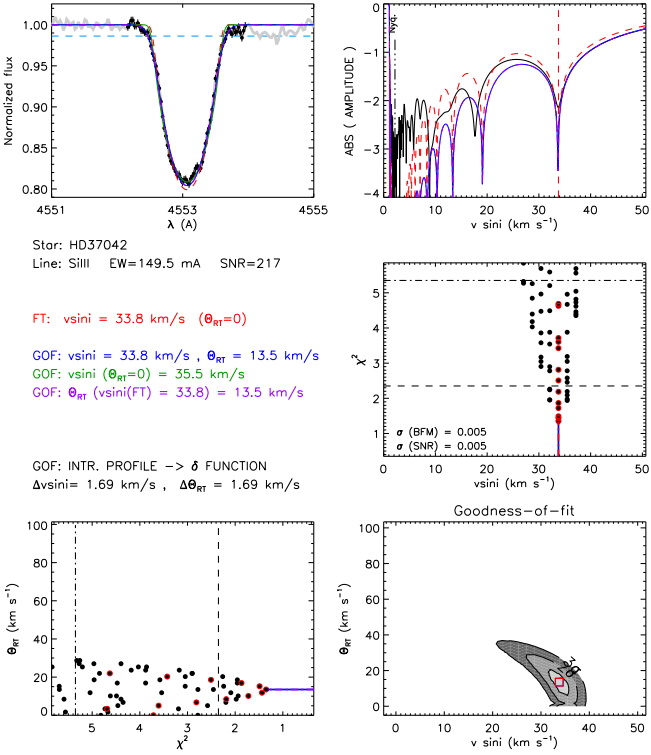}
\caption{Example of graphical output from the {\tt iacob-broad} tool for the case
of the B0.7\,V star HD37042. Five main graphical results are presented: (a) The line 
profile (upper left); (b) the FT of the line (upper right); and (c) 2-D $\chi^2$-distributions 
resulting from the GOF analysis (lower right) and their projections (above and to the left). 
See text for furthest explanations.}
\label{fig1}
\end{figure}

The goodness-of-fit (GOF) technique is based on a simple line-profile fitting in which an 
intrinsic profile is convolved with the various considered/investigated line-broadening profiles 
and compared with the observed profile using a $\chi^2$ formalism. An example of the
application of this technique to a sample of Galactic early-B~Sgs can be found in \cite{Rya02}.
As already pointed out by these authors, one of the caveats of the GOF method to 
disentangle rotational from \macrot\ broadening is the high degree of 
degeneracy in the resulting profiles when convolved with different (\vsini, $v_{\rm m}$) combinations 
(resulting in a banana-shaped $\chi^2$ distribution, see e.g., Figs.\ref{fig1} and \ref{fig2}).
More, as we show in Sect.~\ref{Sect33}, different 
descriptions of the \macrot\ profile lead to different \vmacro\ and \vsini\ measurements.
In addition, the derived \vsini\ and \vmacro\ may be erroneous if the total broadening is 
comparable to the intrinsic broadening of the line and the later is not properly taken into account.
This second warning must be specially kept in mind when using \ion{He}{i-ii} lines
due to the effect of Stark broadening \citep[see e.g.,][]{Ram13}, and in those 
stars having relatively low \vsini\ and \vmacro\ (due to the effect of microturbulence, 
see further discussion in Sect.~\ref{Sect34}).

While the GOF technique allows for the full
characterization of the line-broadening in terms of projected rotational and \macrot\ 
velocities, the identification of the first zero in the FT only provides a measurement of \vsini.
The estimation of the extra-broadening in this case requires a further comparison of synthetic 
profiles broadened by rotation and \macro\ and the observed profiles (either in the 
lambda or Fourier domain). \cite{Gra76, Gra05} proposes the use of the full information 
contained in the FT of the line-profile. The strategy we follow in this paper is based on 
a combined FT+GOF approach.

\subsection{{\tt iacob-broad}: a combined FT+GOF analysis}\label{Sect32}

%
\begin{figure}[t!]
\centering
\includegraphics[width=8.5 cm,angle=0]{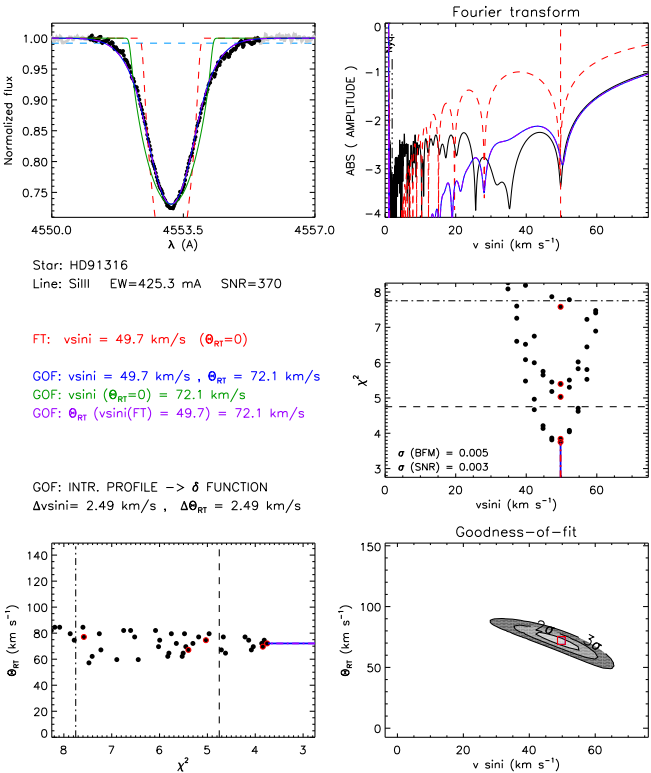}
\caption{As Fig. \ref{fig1} for the case of the B1\,Iab\,N str. star HD\,91316.
}
\label{fig2}
\end{figure}

The general philosophy of the {\tt iacob-broad} IDL procedure is simple: extract
in just one shot as much information as possible about the line-broadening 
characteristics of a given stellar line-profile in a versatile and user friendly way. 
This information is obtained by the combined application of the two methodologies
described in the previous subsection to an observed line-profile.

Before starting the combined FT+GOF analysis, the tool allows for a 
preprocessing of the line profile (e.g., renormalization of the
local continuum, clipping part of the line-profile to eliminate nebular contamination)
and for the selection of the intrinsic profile (either a delta-function or a line profile 
from a stellar atmosphere code). For the purposes of the present work, 
two types of \macrot\ profile are incorporated into the GOF analysis: 
an isotropic Gaussian, and a radial-tangential description \cite[see][for definitions]{Gra05}.

Figs.~\ref{fig1} and \ref{fig2} show the graphical output resulting from the {\tt iacob-broad}
analysis of the \ion{Si}{iii}\,$\lambda$4452 line from the IACOB spectra of HD\,37042 (B0.7\,V) and 
HD\,91316 (B1\,Iab\,N str.). 
Each graphical output includes as much information as possible concerning
the broadening characterization of the studied line. Five main graphical results 
are presented: (a) The line profile (upper left); (b) the FT of the line (upper right); 
and (c) 2-D $\chi^2$-distributions resulting from the GOF analysis (lower right) and 
their projections (above and to the left). In addition, the following
information is provided in the same figure (the associated uncertainties are 
calculated and provided on the screen and output files):
\begin{itemize}
\item the \vsini\ corresponding to the first zero of the FT;
\item the \vsini\ and \macrot\ velocity resulting from the GOF, assuming both are free parameters;
\item the \vsini\ resulting from the GOF, when the \macrot\ velocity is fixed to zero (i.e., this 
result assumes that the full broadening is due to rotation and is directly comparable to 
previous studies not accounting from the extra-broadening);
\item the \macrot\ velocity resulting from the GOF when the \vsini\ is fixed to the value corresponding
to the first zero of the FT.
\end{itemize}
For the sake of simplicity, we denote these different values as \vsini(FT), \vsini(GOF), \vmacro(GOF), 
\vsini(GOF, \vmacro=0), and \vmacro(FT+GOF), respectively, along the paper. We will use \vmacro\ when 
generically referring to \macrot\ velocity. On the other hand, since two different characterizations of 
this broadening agent are used in this paper, namely an isotropic Gaussian profile and a radial-tangential 
profile, we will quote these two possibilities as \TG\ and \TRT, respectively.

For each set of resulting broadening parameters, the initially considered synthetic profile 
is convolved with the (\vsini, \vmacro) pair and degraded to the resolving power of the 
analyzed spectra; then the corresponding profiles and the resulting Fourier transforms are 
overplotted in the various panels using the same color code as in text
for better identification.

The graphical output allows the user to understand the impact of the different 
contributions to the line profile broadening, and evaluate the reliability of the various solutions. 
From the two examples shown in Figs. \ref{fig1} and \ref{fig2}, HD\,37042 represents a typical case in which rotation 
dominates the broadening of the \ion{Si}{iii} line. As a result of this and of the high S/N ratio 
of the line, the broadening analysis results in an excellent agreement between \vsini(FT) and 
\vsini(GOF). In addition, \vsini(GOF, \vmacro=0) is very close to these two values. 
HD\,91316, on the other hand, illustrates the case of a star in which an important (dominant) 
\macrot\ broadening contribution affects the \ion{Si}{iii} line-profile. In this case,
\vsini(FT)\,=\,vsin(GOF), but inspection of the synthetic line profile
and its corresponding FT convolved with this \vsini\ value (the red lines in the upper panels of Fig.~\ref{fig2}) 
already tell us that rotation is not the only broadening agent. As a consequence,
the derived \vsini\ assuming rotation as the only broadening agent results in a too large \vsini\
value. We emphasize here the role played by the spectral resolution. At a resolution 
significantly lower than the present observations (R\,=\,46000) the disentangling of rotation and 
\macrot\ broadenings would be much difficult.

These two cases were deliberately selected as examples in which there is perfect 
agreement between FT and GOF results. However, this will not always be the case. Under certain situations,
\vsini(FT)\,$\neq$\,\vsini(GOF) and the user will have to use the provided information to decide which one 
is the correct solution (if any!). Some of the possible difficulties are presented in the next subsections.

\subsection{Gaussian vs. radial-tangential \macrot\ profile}\label{Sect33}

\begin{figure}[t!]
\centering
\includegraphics[width=8. cm,angle=0]{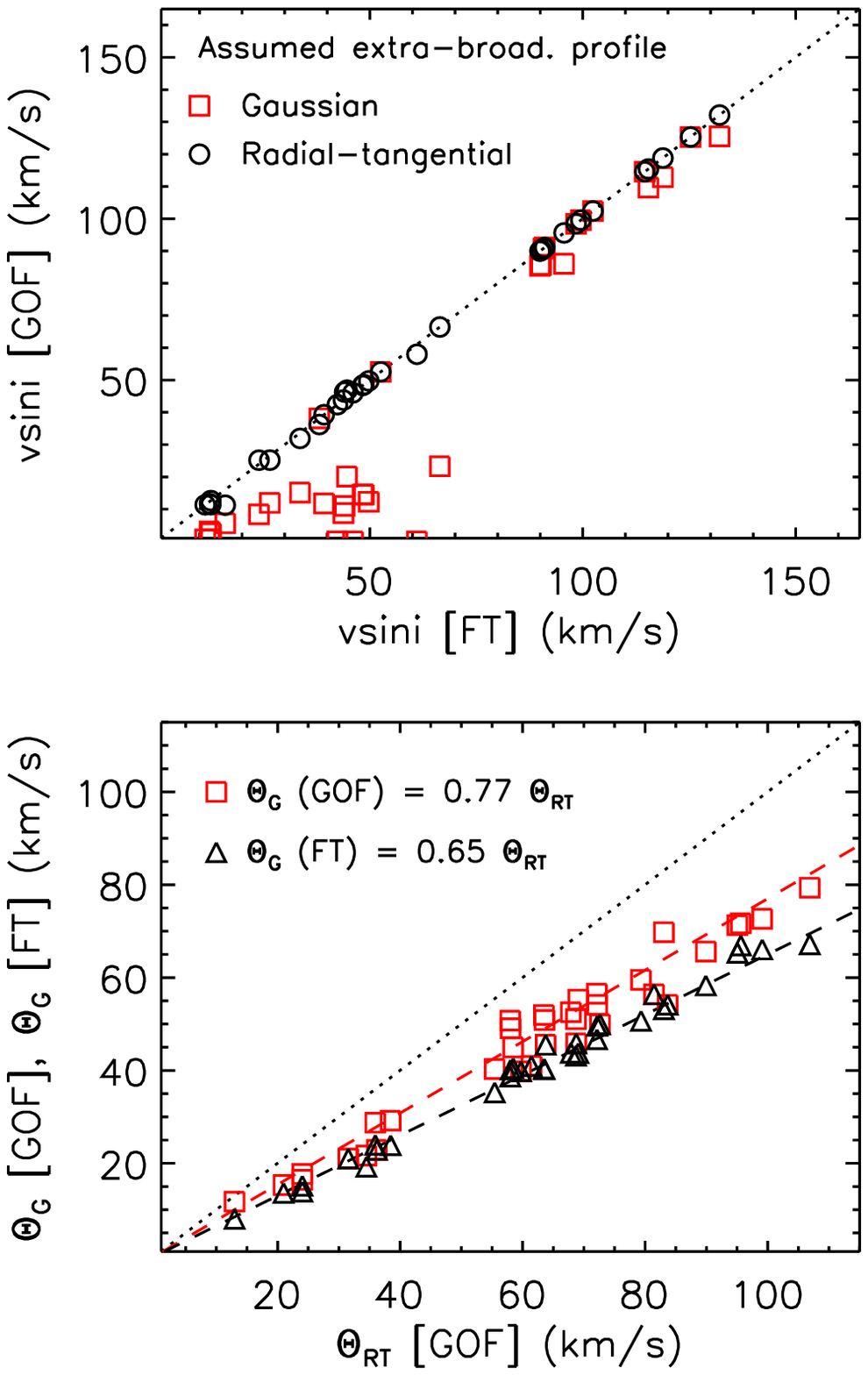}
\caption{Comparison of \vsini\ and \vmacro\ estimates resulting from the {\tt iacob-broad} 
analysis of a subsample of stars \vsini\,$\le$\,150 \kms\ and \TRT\,$\ge$\,0.5\vsini. 
Two different definitions of the \macrot\ profile are considered in the case
of GOF results: a isotropic Gaussian profile and a radial-tangential profile. 
[Upper panel] Comparison of \vsini\ values obtained by means of the FT and GOF
techniques. [Lower panel] Comparison of \macrot\ velocities resulting from the GOF 
analysis. Two options are investigated in the case of the Gaussian profile: 
[red squares] leaving also \vsini\ as a free parameter; [black triangles] fixing \vsini\ 
to the value provided by the FT. Dotted lines (both panels) represent the 
one-to-one relation; black and red dashed lines (bottom panel) correspond to the linear
fit to the data represented as black triangles and red squares, respectively.}
\label{fig3}
\end{figure}

While the presence of a non-negligible non-rotational broadening has been firmly confirmed
(see Sect.~\ref{Sect1}), its physical origin is still a 
puzzling question. As a consequence, we still lack a formal description of the 
corresponding broadening profile. Meanwhile, following
some guidelines developed in the context of cool stars, two different definitions of the 
(up to now called) \macrot\ broadening have been commonly used for OB-type stars 
in the last years: an isotropic Gaussian definition and a radial-tangential definition.

In \cite{Sim10} we pointed out that the values of \vsini\ and \vmacro\ 
determined by means of the GOF method depend on the assumed \macrot\ profile.
There we suggested, based on empirical arguments, that a radial-tangential
prescription is more appropriate results. In this section we re-assess this topic and provide
further evidence that the radial-tangential profile gives more consistent results.

For this investigation we considered a subsample of the IACOB spectra analyzed in this paper, 
fulfilling the following criteria: (a) Well defined zeroes in the FT (b) \vsini\,$\le$\,150 \kms, 
and (c) \vmacro\,$\ge$\,0.5\,\vsini. We carried out the analysis twice, 
first assuming a radial-tangential profile and then an isotropic Gaussian profile, all
other conditions being exactly the same. The upper panel of Fig.~\ref{fig3} compares the 
\vsini(GOF) values obtained in each case with the corresponding \vsini(FT). While there is a perfect 
agreement between \vsini(GOF, \TRT) and \vsini(FT), \vsini(GOF, \TG) values
are systematically too low compared with the FT determinations. The lower panel 
illustrates the fact that the two different descriptions lead also to different values of \vmacro. 
There is a clear correlation between
the \TG\ and \TRT\ measurements, but \TG\ values are systematically lower ($\sim$\,76\%
in the case of free \vsini\ and 65\% in the case of \vsini\ fixed to its FT value).

This result is a consequence of the different characteristics of both
profiles. Basically, the isotropic Gaussian profile results in a broader core than the radial-tangential 
profile for the same wing extension. As a consequence a lower \vsini\ is needed in 
the case of the Gaussian profile. This effect is negligible when \vsini\ is large 
enough to dominate the contribution to the line core broadening
(see, e.g., those cases with \vsini\,$>$\,100~\kms\ in Fig.~\ref{fig3}).

\cite{Rya02} were among the first authors trying to quantify the relative contribution of rotation
and \macrot\ broadening in B\,Sgs. They applied a GOF method with an isotropic Gaussian
definition of the \macrot\ profile, and concluded that a model where \macro\ 
dominates and rotation is negligible is more acceptable than the reverse scenario. 
Their \vsini\ values were, however, too low compared
with results obtained afterwards based on the Fourier transform. This puzzling result has been
neither investigated nor explained in the literature yet. Our investigation definitely 
explains this discrepancy\footnote{We remark here that this effect can be also present in 
some of the more critical cases presented in Figs.~6, 7, and 8 in \cite{Aer09}.} and shows that 
\vmacro\ values from different sources can only be combined with care.

Three main conclusions can be extracted from our investigation:
\begin{itemize}
\item While the physical origin of the \macrot\ broadening is still not known
(and hence we lack a formal calculation of the profile describing the extra-broadening),
the results presented in the upper panel of Fig.~\ref{fig3} support that 
a radial-tangential \macrot\ profile is better suited to determine rotational velocities 
in OB stars (based on a GOF approach) than a Gaussian profile.
\item The excellent agreement between \vsini(FT) and \vsini(GOF, \TRT)
strengthens the suitability of the FT method for the determination of rotational velocities in OB stars.
We also refer the reader to Sects.~\ref{Sect34} and \ref{Sect41} for a further discussion of this statement.
\item When investigating relationships between the size of the extra-broadening and other
stellar parameters, one must combine \macrot\ velocities from the literature with care since there
is a systematic offset between values obtained under the various assumptions usually considered
by different authors.
\end{itemize}

\subsection{The effect of microturbulence}\label{Sect34} 

\begin{figure}[t!]
\centering
\includegraphics[width=12. cm,angle=90]{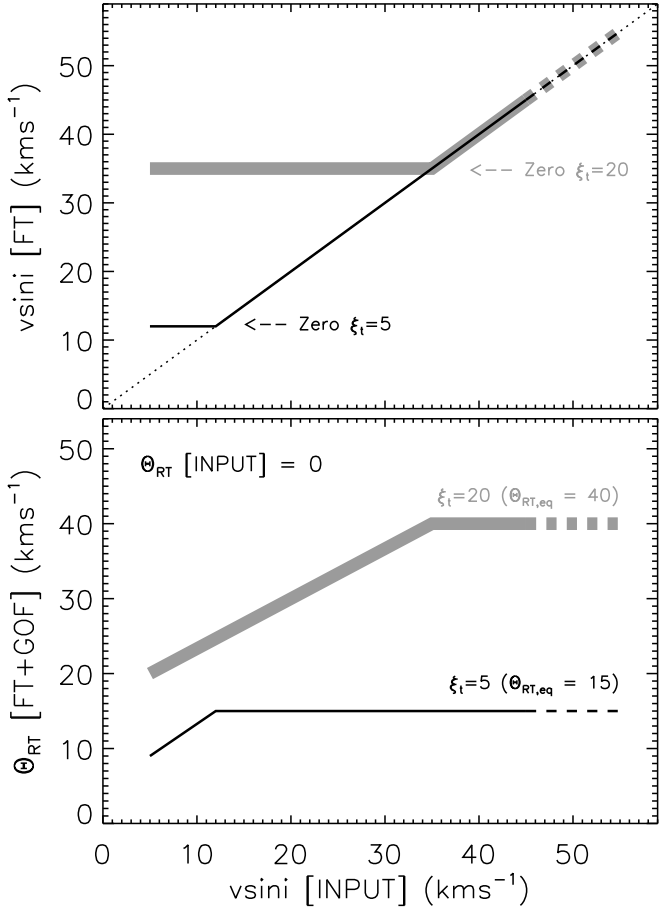}
\caption{Effect of microturbulence on the derived \vsini\ and \TRT\ when
microturbulent broadening is not taken into account in the FT+GOF
analysis. Black and grey curves indicate the derived broadening parameters
resulting from the FT+GOF analysis of a \ion{Si}{iii} line synthesized assuming 
two different values of microturbulence (5 and 20 \kms, respectively). 
Top panel: derived (via FT) vs. input \vsini\ values. The position of the zeros
associated with microturbulence are also indicated. Bottom panel: Derived 
\TRT\ values for the range of considered input \vsini\ values. 
}
\label{fig4}
\end{figure}

\begin{figure}[t!]
\centering
\includegraphics[width=6. cm,angle=90]{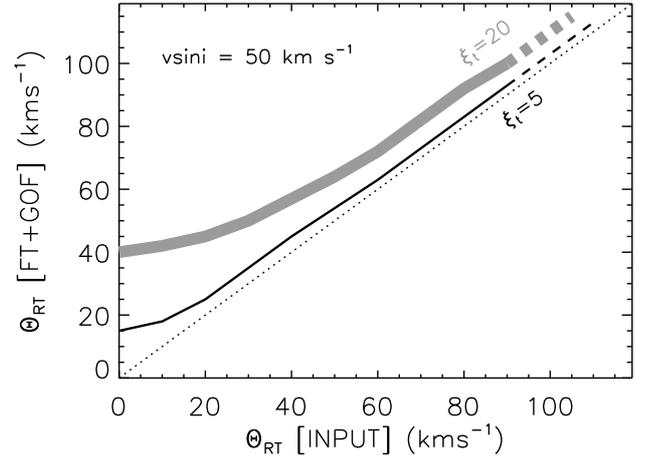}
\caption{Effect of microturbulence on the derived \TRT\ when
microturbulent broadening is not taken into account in the FT+GOF
analysis. Black and grey curves indicate the derived \TRT\
resulting from the FT+GOF analysis of a \ion{Si}{iii} line synthesized 
assuming two different values of microturbulence (5 and 20 \kms, 
respectively) and convolved to a fixed \vsini\,=\,50 \kms, and the
\TRT\ values indicated in the abscissa. 
}
\label{fig5}
\end{figure}

In many cases, one has to perform a line-broadening analysis without having
access to a realistic intrinsic profile (i.e. computed with a stellar
atmosphere code). In that case, it is habitual to consider a delta-function
with the same equivalent width as the observed line. As a consequence, the
GOF analysis incorporates all line-broadening not produced by rotation (and 
spectral resolution) to the \vmacro\ parameter. One can minimize this effect
using metal lines where, contrarily to the case of \ion{H}{} and \ion{He}{i-ii}
lines, the Stark broadening is negligible. However, even in this case there is 
yet an important source of line-broadening which may lead to an overestimation 
of the \macrot\ broadening: microturbulence.

Microturbulence also has an effect when performing an FT analysis
to determine \vsini. As pointed out by \cite{Gra73}, 
microturbulence also produces zeroes in the Fourier transform at (high) frequencies 
associated with low values of \vsini\ that may be wrongly identified  
as the zeroes associated with the rotational broadening below a certain \vsini\ limit. 

In \cite{Sim07} we presented a few notes about the effect of microturbulence on the determination 
of \vsini\ via the FT method for the case of OB-type stars. In this section, we extend further that
study and also investigate the impact that microturbulence can have on our \TRT\ 
measurements. To this aim, we performed a couple of exercises in which a set of synthetic 
\ion{Si}{iii}\,$\lambda$4552 lines computed with the stellar atmosphere code FASTWIND 
\citep{San97, Pul05} were analyzed using the 
{\tt iacob-broad} procedure in the same way as the observed spectra. 
Two values of microturbulence (5 and 20 \kms) were
considered as representative for Dwarfs and Supergiants, respectively.
In a first exercise, each of these two profiles was convolved with different values of 
\vsini\ ranging from 5 to 400 \kms, and no \macrot\ broadening was incorporated to 
the synthetic lines. In a second exercise, \vsini\ was fixed to 50 \kms\ and a 
\macro\ (\TRT) ranging from 10 to 100 \kms\ was added to each profile. 

Results from these exercises are presented in Figs.~\ref{fig4}~and~\ref{fig5}, were 
the derived values of \vsini(FT) and \TRT(FT+GOF) are compared to the input values. Top panel 
in Fig.~\ref{fig4} indicates that there is perfect agreement between the derived and 
the input \vsini\ when the first zero associated with
rotation is located at lower frequencies (larger \vsini) than the zero resulting from
microturbulence. On the other hand, for lower \vsini\ values the later would be 
wrongly attributed to rotation and hence the plateau in the derived \vsini. Therefore, 
microturbulence imposes an upper limit below which \vsini\ cannot be correctly derived 
by means of the first zero in the FT. It is important to note that this \vsini\ limit 
depends on the value of the microturbulence and the equivalent width of the line. The 
larger these parameters, the higher the limit on the detectable \vsini. In the case of the 
example considered in Fig.~\ref{fig4}, this limit can be up to 35~\kms\ for the case 
of \vmicro\,=\,20~\kms.

Interestingly, the derived \vsini(GOF) values agree with \vsini(FT) determinations for all 
the cases considered in these exercises (even when they are wrong!). This seems to indicate 
that in the low \vsini\ regime, when microturbulent broadening dominates, the profile
associated to microturbulence (at least in the form resulting from FASTWIND computations) 
can be mimicked by a rotational profile with a given \vsini\ (plus the inclusion of some 
extra-broadening to fit the extended wings on the line).

The lower panel of Fig.~\ref{fig4}, along with Fig.~\ref{fig5} illustrate that, as expected, a simple 
\vsini\,+\,\TRT\ analysis without accounting for the effect of microturbulence on 
the line-profile systematically leads to an overestimation of \TRT. This effect is more important
when the relative contribution of microturbulence vs. \macro\ is larger. In the example 
considered in bottom panel of Fig. \ref{fig4}, a microturbulence 
of 5/20 \kms\ in a star without any \macrot\ broadening contribution, could be erroneously 
interpreted as having \TRT\,$\sim$\,15/40 \kms, respectively. As indicated in Fig.~\ref{fig5} the 
relative overestimation becomes smaller with increasing \macro. Interestingly, the input 
\macro\ is not perfectly recovered even at large values of \TRT.\\
\newline
Two main conclusions can be highlighted from our study. If the effects of microturbulence are not
taken into account: 
\begin{itemize}
\item[a)] Any \vsini\ measurement (either from FT or GOF) below $\sim$\,40~\kms\ must be considered 
as an upper limit (specially in the case of O and early-B~Sgs, where the large values of microturbulence 
are commonly found); 
\item[b)] Any 
\TRT(GOF) measurement must be considered as an upper limit of the actual \macrot\ broadening. 
In particular if \TRT\,$\leq$\,15/40 \kms\ in dwarfs/supergiants, the derived \macrot\ 
broadening may be actually produced by microturbulence; on the other hand, even \TRT\ values 
above 25/60 \kms\ may be actually smaller by 5/15 \kms. 
\end{itemize}

We should note that all the numbers indicated above must be considered as illustrative since 
the actual values depend of the equivalent width of the investigated line, and the specific
microturbulence derived for the star under study. 

The two exercises presented in this section warn us about the limitations of the proposed FT+GOF
strategy. These warnings are specially important when interpreting results in the low 
\vsini\ and \TRT\ regimes. Also, when the derived line-broadening 
parameters (specially \TRT) are used for the spectroscopic determination of the stellar parameters and/or 
abundances by fitting metal lines. In that case, these values will have to be corrected downwards depending 
on the derived/assumed microturbulence.


Some of these limitations may be surpassed if lines with different equivalent widths are 
analyzed (for the case of \vsini), more zeroes are considered in the FT,
and synthetic lines incorporating the microturbulence resulting
from a quantitative spectroscopic analysis are considered as initial profiles in the fitting procedure
(for the case of \TRT). We plan to explore these ways for future improvements.

\subsection{Applicability and limitations in the IACOB sample}\label{Sect35}

Before presenting the results of the line-broadening analysis of the whole sample of Galactic OB stars 
considered in this study (Sect.~\ref{Sect4}), we summarize below some important points (apart from
those more general ones indicated in previous sections) which must be reminded to correctly 
interpret the results from the {\tt iacob-broad} analysis of the IACOB spectra.

\subsubsection{Resolution and spectral sampling}\label{Sect351} 
The resolving power of the IACOB spectra is 46000 and 25000 (FIES medium and low resolution modes, 
respectively), both having a spectral dispersion of 0.025 \AA/pix.
The lower \vsini\ limit detectable via FT \citep[related to the Nyquist frequency, see][]{Sim07}
is $\sim$\,2\,\kms, and the velocity corresponding to the spectral resolutions is $\sim$\,5 
and $\sim$10 \kms\, respectively.


\subsubsection{Signal-to-noise ratio}\label{Sect353} 

The S/N ratio of the continuum adjacent to the analyzed lines is normally above 150. There are, however, 
a few cases with a S/N ratio slightly lower than 100. 

To investigate the effect of noise for the specific case of the spectroscopic dataset analyzed
in this study, we obtained 13 spectra of HD\,91316 (one of the examples presented in 
Sect.~\ref{Sect3}, see also Fig.~\ref{fig2}) with varying exposure times leading to S/N ratios 
ranging from 90 to 300. The corresponding 
\ion{Si}{iii}\,$\lambda$4552 lines were analyzed with the {\tt iacob-broad} procedure using the same 
wavelength limits. We found that the derived \vsini(FT) and \vsini(GOF, \TRT) values range between 
36\,--\,53, and 41\,--\,51 \kms, respectively, without any clear correlation with the S/N of the 
continuum. The mean values and standard deviations in each case are 46\,$\pm$\,5 and 46\,$\pm$\,3 \kms,
respectively. Therefore, from this single experiment we can conclude
that, for this spectroscopic dataset, the effect of noise on the derived \vsini\ and \TRT\ does not
seem more relevant than other sources of uncertainty (e.g., local renormalization, 
asymmetry of the line, non detected spectroscopic binarity).

\subsubsection{Broad line stars}\label{Sect354} 

\begin{figure}[t!]
\centering
\includegraphics[width=8.5 cm,angle=0]{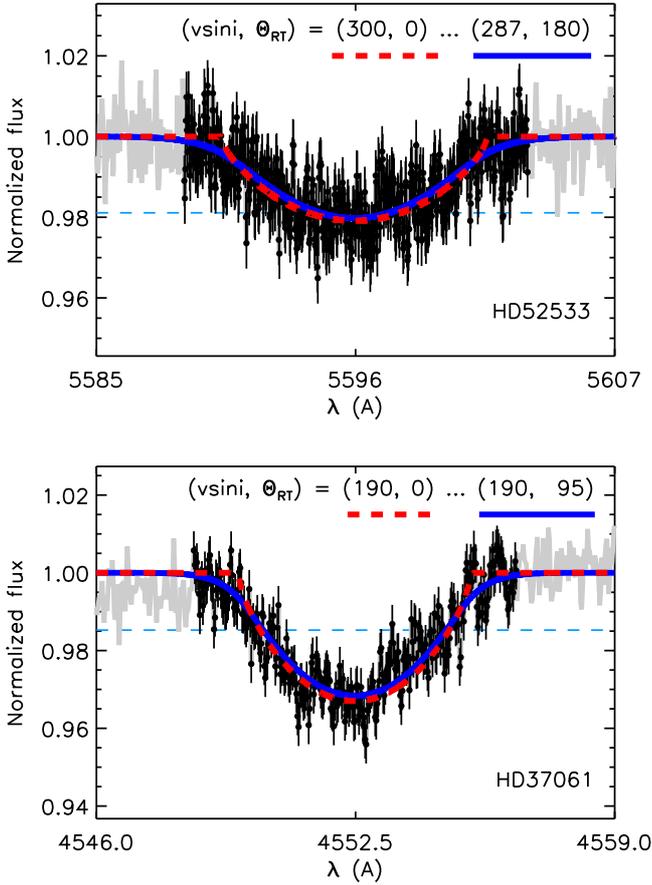}
\caption{A couple of representative examples of broad line stars. In both 
cases the noise makes the accurate measurement of \TRT\ difficult and,
in fact, a model with \TRT\,=\,0 could also perfectly fit the line. Note
that in the case of HD\,52533, the increased extra-broadening forces \vsini\
to be lower than \vsini(FT).}
\label{fig6}
\end{figure}

Stars with projected rotational velocities above $\sim$\,150\,\kms\ deserve a separated section, since
they present some issues not equally important in relatively narrow line stars. The main ones are 
summarized below:
\begin{itemize}
\item The first issue is the possibility of non-detected spectroscopic binaries/multiple 
systems, specially when the various components have large \vsini\ and the radial velocity amplitude
of the system is small.
\item Lines can be really shallow in very fast rotators, and the
effect of noise, blending, and specially continuum normalization becomes important. 
Above a certain \vsini\ value the effect of the extra-broadening only result in
a subtle shaping of the extended wings (see e.g., Fig.~\ref{fig6}), and hence 
a reliable measurement of \TRT\ becomes difficult due to the effect of noise 
and local continuum normalization on the fitting process. We have found from our 
analysis that in stars with \vsini\,$\geq$\,180 \kms\ the values of \TRT\ resulting from
the {\tt iacob-broad} analysis must be considered as upper limits.
\item Equatorial gravity darkening will produce an obscuration of the profile wings associated with the 
larger velocities and thus an underestimation of the actual \vsini\ \citep{Tow04, Fre05}.
While in this paper we present \vsini\ measurements 
for stars above 200 \kms\ obtained in the same way as the global sample and without incorporating 
any correction, we plan to investigate this type of effects in a future paper, also including
\vsini\ measurements provided by the {\em stronger} \ion{He}{i-ii} lines.
\end{itemize}

\section{Analysis of a sample of Northern Galactic OB-type stars}\label{Sect4}

Tables \ref{taba1}\,--\,\ref{taba5} summarize the results from the line-broadening characterization
of the complete sample. The {\tt iacob-broad} procedure was applied to the 
\ion{O}{iii}\,$\lambda$5591 or \ion{Si}{iii}\,$\lambda$4552 lines
depending of the spectral type and luminosity class\footnote{There is a small subsample of 
O9-B0 stars in which the two lines are available. We have found a very good agreement between results 
from the {\tt iacob-broad} analysis of both lines.}, and the derived \vsini(FT), \vsini(GOF), and
\TRT(GOF) are indicated in columns 7\,--\,9. The tables also include information about
the spectral type and luminosity class of the stars (columns 2 and 3), the used line
and its equivalent width (column 4 and 5) and S/N ratio of the adjacent continuum.

Following the results of the exercise presented in Sect.~\ref{Sect33} we assumed a radial-tangential
definition of the \macrot\ profile. Also, it is important to remark that
we assumed a delta-function as intrinsic profile\footnote{A quantitative spectroscopic analysis 
aimed at the determination of the stellar and wind parameters of the whole IACOB sample 
is now in progress. Results will be presented in a forthcoming paper. 
}; therefore, the derived \macro\ includes all sources of non-rotational broadening 
(i.e. microturbulence and any other possible unknown source of broadening). This may have important 
consequences for those cases in which a low value of \vsini\ and/or \macro\ is obtained (see notes 
in Sect.~\ref{Sect34}).

\subsection{\vsini(FT) vs. \vsini(GOF)}\label{Sect41}

\begin{figure}[t!]
\centering
\includegraphics[width=8.5 cm,angle=0]{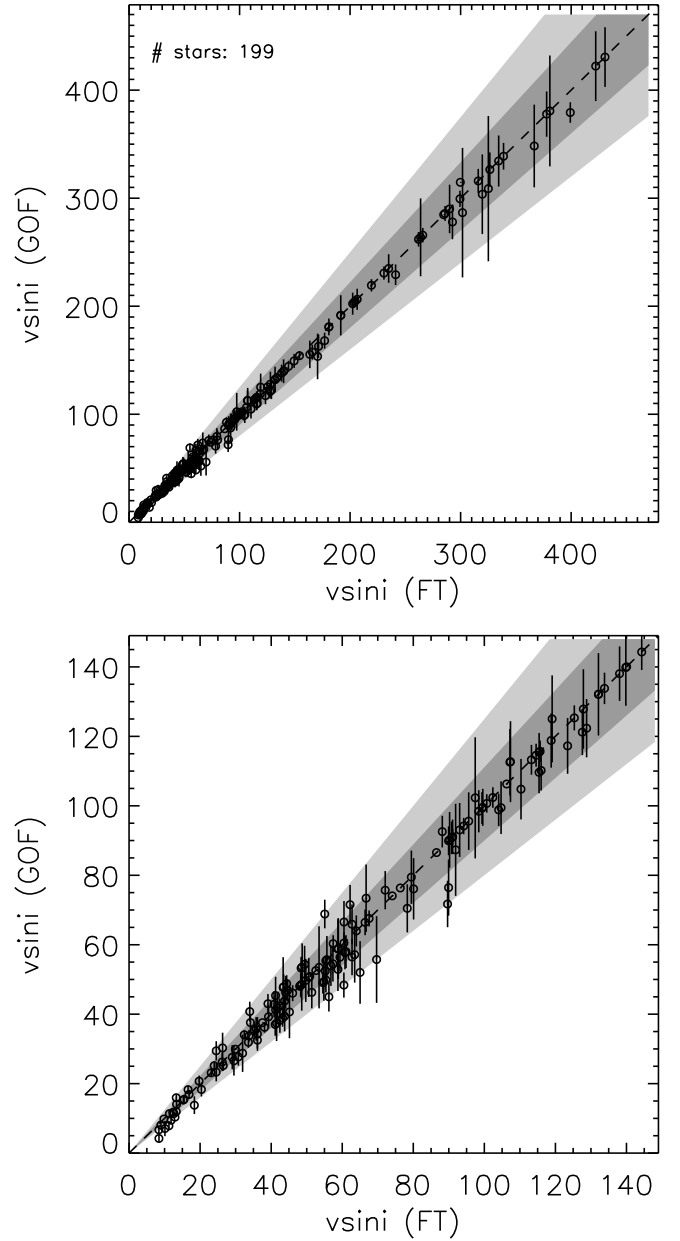}
\caption{Comparison of \vsini\ measurements derived by means of the FT and GOF 
techniques. Upper panel: Full \vsini\ range. Lower panel: Zoom on the low 
\vsini\ range.
}
\label{fig8}
\end{figure}

\begin{figure}[t!]
\centering
\includegraphics[width=8.5 cm,angle=0]{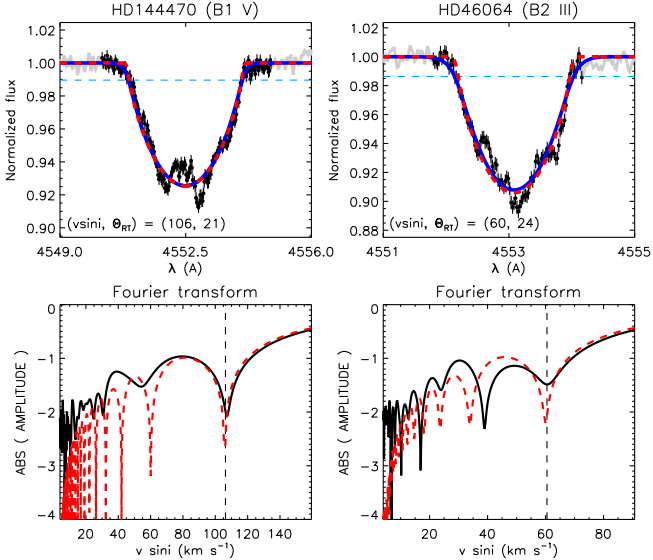}
\caption{A couple of representative examples of the FT+GOF analysis for 
stars showing features associated to $\beta$\,Cep-type non-radial pulsations.
Red dashed lines corresponds to profiles with \vsini(FT) and \TRT\,=\,0. The
blue lines (only indicated in the upper panels) also include the extra-broadening 
resulting from the GOF.}
\label{fig7}
\end{figure}

\begin{figure*}[t!]
\centering
\includegraphics[width=11. cm,angle=90]{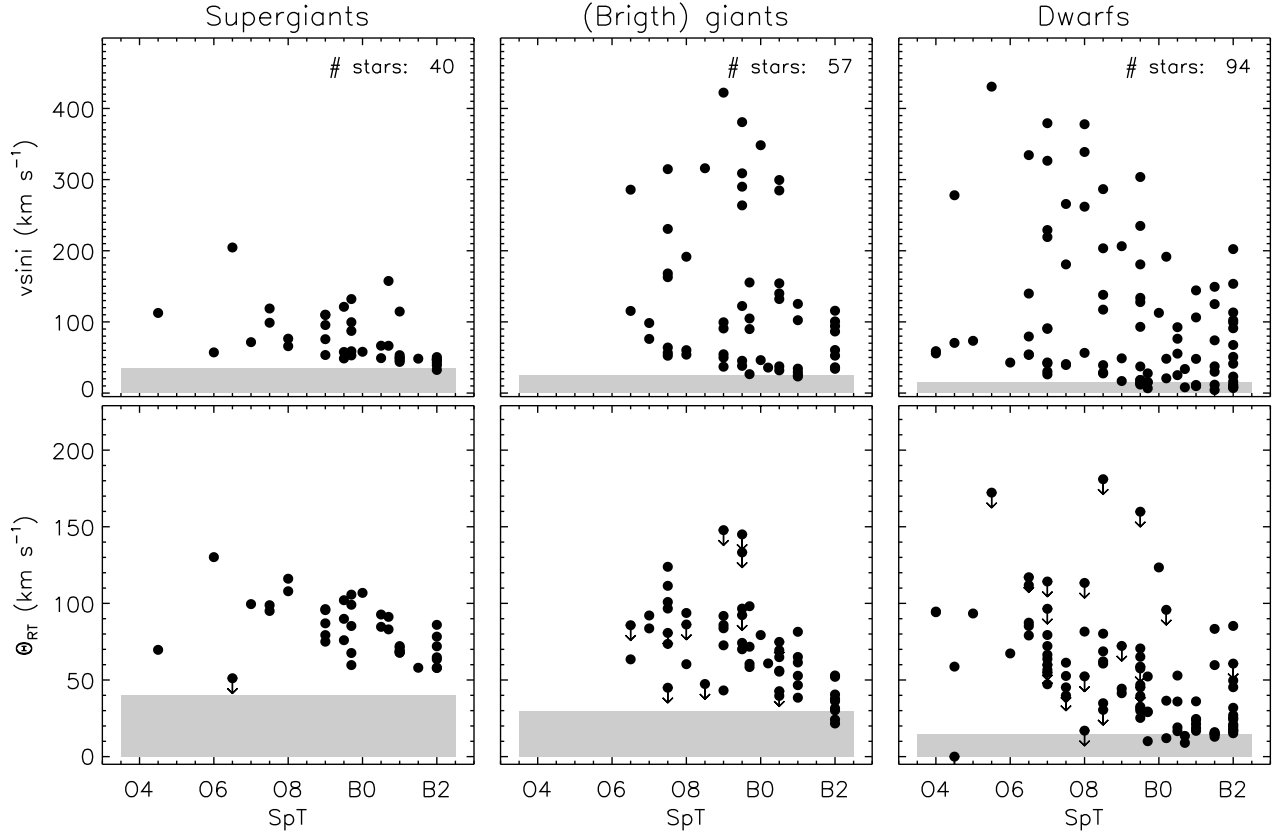}
\caption{\vsini\ and \TRT\ measurements vs. spectral types from the analyzed
sample of OB-type stars. \TRT\ measurements for stars with \vsini\,$\ge$\,180 \kms\
can be only considered as upper limits (see text for explanation) and hence
a downward arrow is added to the corresponding filled circle in lower panels.
}
\label{fig9}
\end{figure*}

Fig.~\ref{fig8} compares the \vsini\ measurements obtained by means of the FT and GOF 
approaches. The one-to-one relation is indicated with a dashed line along with the 10 and 20\%
difference regions in dark and light grey, respectively. Upper panel includes all cases,
while the lower panel zooms the region with \vsini(FT)\,$\le$\,150 \kms.
Both \vsini\ measurements are in very good agreement ($<$\,10\%) for all stars with
\vsini\,$\ge$\,100 \kms. The situation becomes slightly worse for stars with projected rotational
velocities below this limit. The agreement is, however, always better than 20\%, except
for a few cases with \vsini\ measurements below 20\,\kms. Taking into account the results 
of the exercises presented in Sect.~\ref{Sect34}, all \vsini\ values below $\sim$\,15\,--\,35 
\kms\ must be considered as upper limits (specially in the case of Sgs); hence, those cases 
which disagree in more than 20\% in this low \vsini\ region can be naturally explained as
due to the limitations of our methodology. 

Only a small fraction of objects (7 out of the 200 analyzed stars with \vsini\,$\ge$\,35 \kms) 
present differences between \vsini(GOF) and \vsini(FT) in the range 10 -- 20\%. We have explored 
these cases in more detail and found that most of them correspond to lines affected by an important 
\macrot\ broadening contribution (some of them also being slightly asymmetric) in which 
the first zero of the FT is not sharply defined. Although we cannot exclude other possible 
effects, the combination of large \TRT\ plus a not high enough S/N ratio can play an important 
role in the \vsini(FT) determination. As a consequence, most of these cases result in \vsini(FT)
slightly larger than \vsini(GOF, \TRT).

The very good agreement in the derived projected rotational velocities obtained by means of
two independent methods can be considered as an observational evidence of the strength of 
both the FT and GOF methodologies. In particular, it allows us to reassess the reliability 
of the FT method after the word of caution indicated by \cite{Aer09}. Although we cannot 
completely discard that the effect described by these authors (see also Sect. \ref{Sect31})
is not present in our analysis, a detailed inspection of the global graphical output provided 
by the {\tt iacob-broad} tool (see Figs. \ref{fig1}
and \ref{fig2}) --- including the observed and simulated line-profiles and their Fourier 
transforms --- allows us to ascertain that the most critical cases found by 
\citeauthor{Aer09} are not actually represented in the analyzed sample.

A representative example of results from the analysis of a star with a dominant 
macroturbulent broadening contribution is presented in Fig.~\ref{fig2} and 
commented in Sect. \ref{Sect32}. Fig.~\ref{fig7} shows two examples of line 
profiles affected by non-radial pulsations ($\beta$ Cep stars in this specific case). 
Although the profiles are clearly asymmetric and with strong local jumps, their 
analysis does not offer any particular difficulty compared to similar stars
showing smoother profiles (see e.g., Fig.~\ref{fig1}). 

In addition, the good agreement found between \vsini(FT) and \vsini(GOF,\TRT) from the 
analysis of our high resolution, high S/N spectroscopic dataset suggests that {\em for 
worse quality spectroscopic datasets (e.g. fainter Galactic OB stars, or extra-galactic 
objects) we can adopt the values derived from the GOF method, as it is less sensitive to 
noise than the FT method, as far a radial-tangential formulation is adopted} and a 
photospheric, unblended metal line is used for the analysis.

\subsection{Macroturbulent broadening in Galactic OB stars}\label{Sect42}

Fig.~\ref{fig9} shows the \vsini\ and \TRT\ measurements of the whole sample plotted 
against spectral types for three different luminosity classes. Following Sect.~\ref{Sect354}, 
all \TRT\ measurements for stars with \vsini\,$\ge$\,180 \kms\ have been marked as
upper limits in lower panels. We also indicate with grey zones the regions in which 
\vsini\ and \TRT\ measurements can be importantly affected be the effect of microturbulence 
(see Sect.~\ref{Sect34}). In particular, we remind here that, within these regions, any \TRT\ 
measurement could be completely attributable to microturbulence, and any \vsini\
measurement must be considered as an upper limit. 

At this point, the upper panels in Fig.~\ref{fig9} are only included for comparative purposes
(in terms of range of measured values, trends and dispersion) with the information provided
in lower panels. A more detailed discussion of the \vsini\ distributions (in a global
sense and separated by spectral types and luminosity classes) is presented in Sects. 
\ref{Sect432} and \ref{Sect433}.
\begin{figure}[t!]
\centering
\includegraphics[width=8. cm,angle=0]{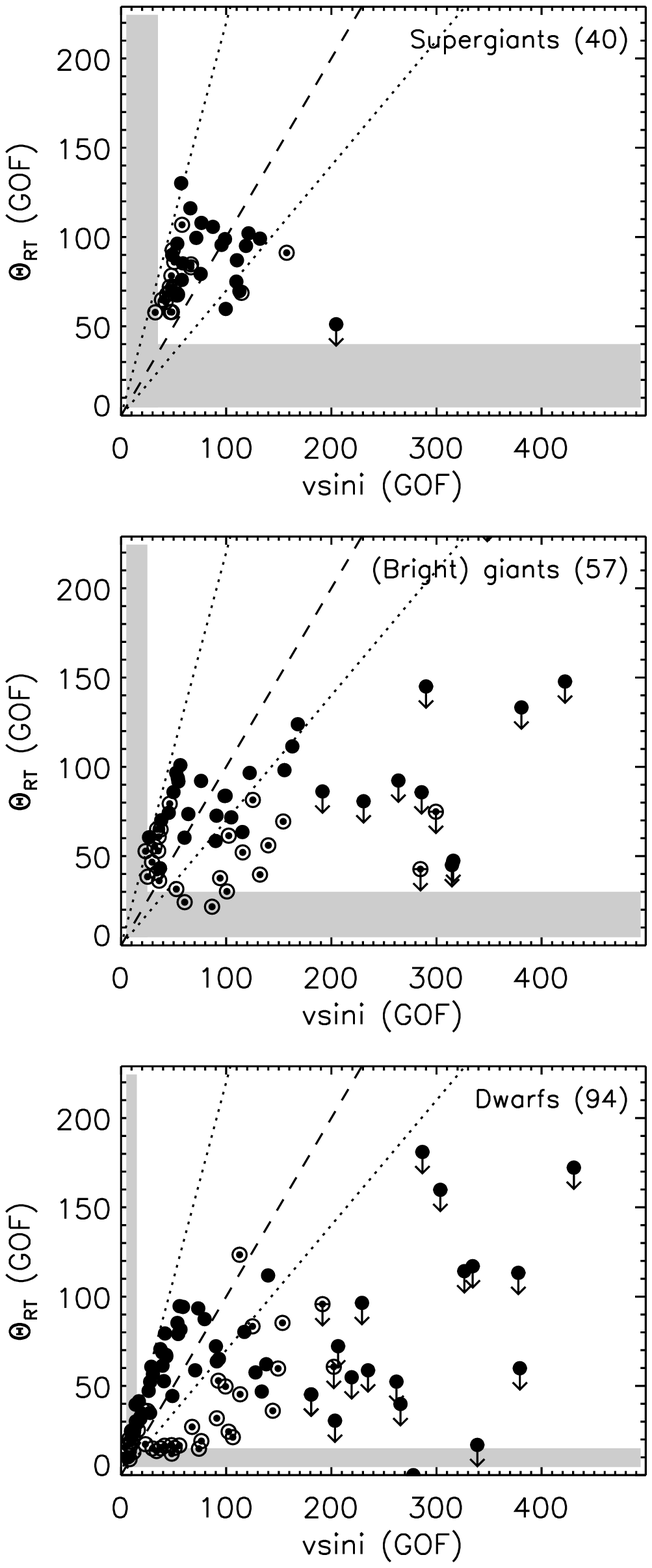}
\caption{\TRT\ vs. \vsini\ for the stars in which there is an agreement between
\vsini(FT) and \vsini(GOF) better than 20\%. Filled and dotted circles indicate
the O and B-type stars, respectively. The indicated values of \TRT\ for all 
those stars with \vsini\,$\ge$\,180 \kms\ must be considered as upper limits.
The regions in which \vsini\ and \TRT\ measurements can be affected by the
effect of microturbulent broadening (see text for explanation) are indicated 
in grey. Dashed and dotted lines indicates the 1:1 and 2.2:1 relations, 
respectively.
}
\label{fig10}
\end{figure}

An important outcome from our study is the confirmation that the so-called \macrot\ 
broadening is present not only in B\,Sgs, but also in O-type stars of all luminosity classes 
\citep[as the analysis of smaller samples have also began to show e.g.][]{Bou12, Mar13}. 
This result has very important implications for our knowledge about 
rotational properties of massive stars based on results from previous studies using methods 
that do not properly disentangle rotation from \macrot\ broadening (see Sect.~\ref{Sect43}). 

It is also interesting to remark the strong correlations (with some dispersion) found 
between \TRT\ and spectral type, and the fact that the size of the \macrot\ broadening 
seems to be slightly larger for Supergiants than for Giants and Dwarfs with the same 
spectral type\footnote{This second statement must be considered carefully, since our
\TRT\ measurements include both the effect of macro- and microturbulence (see notes in Sect.~\ref{Sect34}).}.

Further information about the observational characteristics of the \macrot\ 
broadening (this time vs. \vsini) is presented in
Fig.~\ref{fig10}. Results for the global sample are separated in three panels,
from top to bottom, supergiants, (bright) giants, and dwarfs. O and B-type stars
are presented as filled and dotted circles, respectively. All \TRT\ measurements
for stars with \vsini\,$\ge$\,180 \kms\ are indicated as upper limits. These
are included for completeness, but eliminated from the discussion below.
The regions where \vsini\ and \TRT\ measurements can be affected by microturbulence 
are also indicated. Dashed lines show the 1:1 relation. These lines roughly separate
the regions where either rotational or \macrot\ broadening dominates.

The first thing to note is the very different dependence of 
the \macrot\ broadening with \vsini\ for the three luminosity classes
in O and early-B type stars. While the distribution of stars in the
\TRT\ vs. \vsini\ plane is very similar for O~Supergiants, Giants,
and Dwarfs (with \vsini\,$\le$\,180 \kms!), the relative number of early B-type
stars in which \macro\ dominates decreases from Supergiants to 
Dwarfs. In particular, we do not find any early-B dwarf in our sample with a dominant 
\macrot\ component and, on the contrary, all but two B~Sgs are dominated by  
non-rotational broadening. The two exceptions are HD\,47240
and HD\,191877, both classified as B1\,Ib, and the first one showing disc-like features 
in its spectrum (see also Lefever et al. 2007). We note that these two stars are also the 
only B\,Sgs in our sample having a projected rotational velocity larger than 100~\kms.

Fig.~\ref{fig10} also shows a strong correlation between \TRT\ and \vsini\ 
in the region where \macrot\ broadening dominates. This correlation has 
been also pointed out by \cite{Mar13} in a parallel work combining 
own results and values from other authors in the literature. The correlation occurs for a range in 
the \TRT/\vsini\ ratio between 1 and $\sim$\,2.2 (the later being marked with the dotted line). 
Interestingly, while the relative number of early B-type stars within this region 
diminishes when moving from luminosity class I to V, all O and early-B stars having 
\TRT\,$\ge$\,\vsini\ follow a similar behavior in terms of relative broadening. 

This is not the case in the region below the 1:1 relation, where the line-broadening 
behavior of O and early-B stars is very different. On the one hand, O-type stars 
mainly concentrate on a second sequence characterized by having \TRT/\vsini\,$\sim$\,0.7 
and which is independent of luminosity class. On the other hand, in the case of early
B giants and dwarfs (only two Sgs are found in this region), the \macrot\ broadening 
contribution is negligible (and probably attributable to microturbulence, see below)
below \vsini\,$\sim$\,80\,--\,100 \kms, and seems to increase with \vsini\ above this
value.

It is also interesting to have a look at the percentage of stars found within 
or close to the region where the measured \TRT\ could be interpreted as microturbulence
(see also Fig. \ref{fig9}). All supergiants (but one with \vsini\,$\sim$\,200) have 
\TRT\ values above the 40 \kms\ limit indicated in Sect.~\ref{Sect34}. In the case
of (bright) giants there are a few early-B stars (but no O-type stars) close to the 
limit of 30 \kms. Interestingly, some of them correspond to non-radial $\beta$\,Cep-type 
pulsators (the case illustrated in Fig. \ref{fig7}). Finally, the number of dwarfs with 
\vsini\,$\le$\,150 \kms\ in which the measured \TRT\ can be actually attributed to microturbulence is larger. 
These are mainly early-B and late-O dwarfs.   

These results impose strong observational constraints to any
theoretical attempt to provide an explanation for this line-broadening
of still not confirmed physical origin. While in this paper we mainly concentrate on
the impact of \macrot\ broadening on measurements of projected
rotational velocities in Galactic OB stars, in a forthcoming paper 
\citep[see][for some preliminary results]{Sim12} 
we will investigate whether the pulsational hypothesis proposed by \cite{Luc76} and 
\cite{Aer09} fulfills the above mentioned observational requirements. We 
also refer the reader to \cite{Mar13} where some first steps towards the 
understanding of the observational characteristics of the extra line-broadening as 
a function of stellar parameters and evolution is presented.

\subsection{Projected rotational velocities in Galactic OB stars}\label{Sect43}

\subsubsection{Comparison with previous works: the effect of \macrot\ broadening}\label{Sect431}

The main reference studies of rotational velocities in large samples of 
Galactic O and early B-type stars are those by \cite{Sle56}, \citeauthor{Con77} (1977, CE77), 
\citeauthor{Pen96} (1996, P96) and \citeauthor{How97} (1997, H97). All these studies 
based their \vsini\ measurements 
in methodologies that assume that the line profile is exclusively broadened by rotation. 
These and even earlier authors \citep[viz.][]{Str52} already indicated, 
following indirect statistical arguments, that rotation was very likely not the 
only broadening agent of metal lines in early-type stars. They argued that 
the absence of very narrow line stars in their large samples could be explained 
by assuming the presence of a type of extra line-broadening.

As stated in the introduction, the advent of high resolution, very efficient 
spectrographs has allowed us to surpass the limitations imposed by observations 
in the first attempts to separate the various broadening contributions \citep[e.g.][]{Sle56}, 
as well as to confirm their suspicion. The analysis of these improved datasets 
definitely shows that \vsini\ measurements not accounting for the extra-broadening 
overestimate the actual projected rotational velocities 
\citep[see][and references therein]{Sim11} 

Comparisons with earlier studies not considering \macro\ can be found, among others, 
in \cite{Fra10}, who presented a comparison of 42 B\,Sgs in common with with H97 (with 
spectral types ranging from B0 to B5), or in a study parallel to ours by \cite{Mar13}, 
who have compared their \vsini(FT) results with those obtained by H97 for a 
small (but representative) sample of Galactic O-type stars. In this paper, we extend 
such an investigation using a larger sample of Galactic O and early-B type 
stars analyzed homogeneously.

For the sake of clarity, and taking into account that similar conclusions are
expected to arise when considering the other studies, we concentrate
our investigation in the comparison of our measurements with those derived by H97. 
These authors provided estimates for the $v_{\rm e}$\,sin$i$ linewidth parameter
for 373 Galactic O-type and early B-Sgs obtained by means of the application of a 
cross-correlation technique to high dispersion IUE spectra. The derived quantities
are hence representative of the projected rotational velocities which will be 
derived when only rotation is taken into account (i.e. are similar to the cases of
Slettebak, Conti \& Ebbets, and Penny). \\

\begin{figure*}[t!]
\centering
\includegraphics[width=5.8 cm,angle=90]{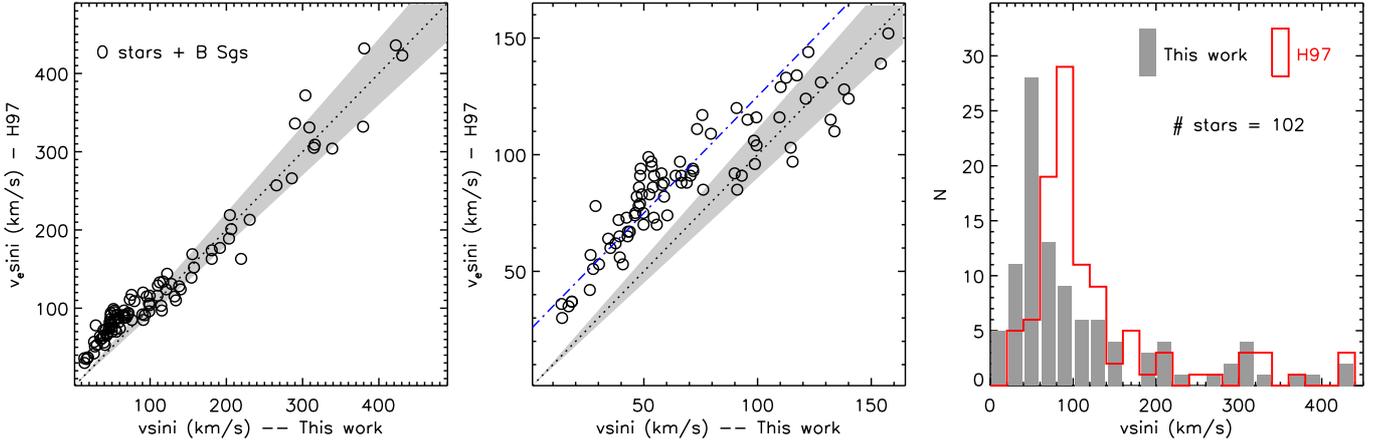}
\caption{Comparison with H97 (all stars in common). Leftmost panel: 
$v_{\rm e}$\,sin$i$ (H97) vs. \vsini(GOF); the one-to-one relation is indicated with 
a dotted line and the $<$~10\% difference region in grey. Center panel:
same as before but zoomed in the region below 150 \kms\ (dash-dotted line
corresponds to \vsini(GOF)\,+\,28~\kms). Comparison of the corresponding
histograms (bin size = 20 \kms) resulting from both studies.
}
\label{fig11}
\end{figure*}

We have $\sim$\,100 stars in common with them (namely $\sim$\,80 O-type stars and 
$\sim$\,20 early-B\,Sgs). Results from the comparison of the global
sample are presented in Fig.~\ref{fig11}. As expected, our estimates are systematically
lower below \vsini\,$\sim$\,100 \kms. Globally, there is a systematic offset in
this \vsini\ regime of $\sim$\,25 \kms; however, differences of up to $\sim$\,50 \kms\ 
can be found in some cases with intermediate projected rotational velocities. 
On the other hand, while there is a relatively good agreement above \vsini\,$\sim$\,120 \kms, 
the intermediate region presents a combined situation in which the extra-broadening 
is still affecting the \vsini\ measurements in some cases (mainly O stars, see below).
The impact on the global \vsini\ distribution is clear: the peak of the distribution
is shifted downwards from $\sim$\,90 \kms\ to $\sim$\,50 \kms, and the number of cases with \vsini\
below 40 \kms\ is multiplied by a factor 2.

\begin{figure*}[t!]
\centering
\includegraphics[width=5.8 cm,angle=90]{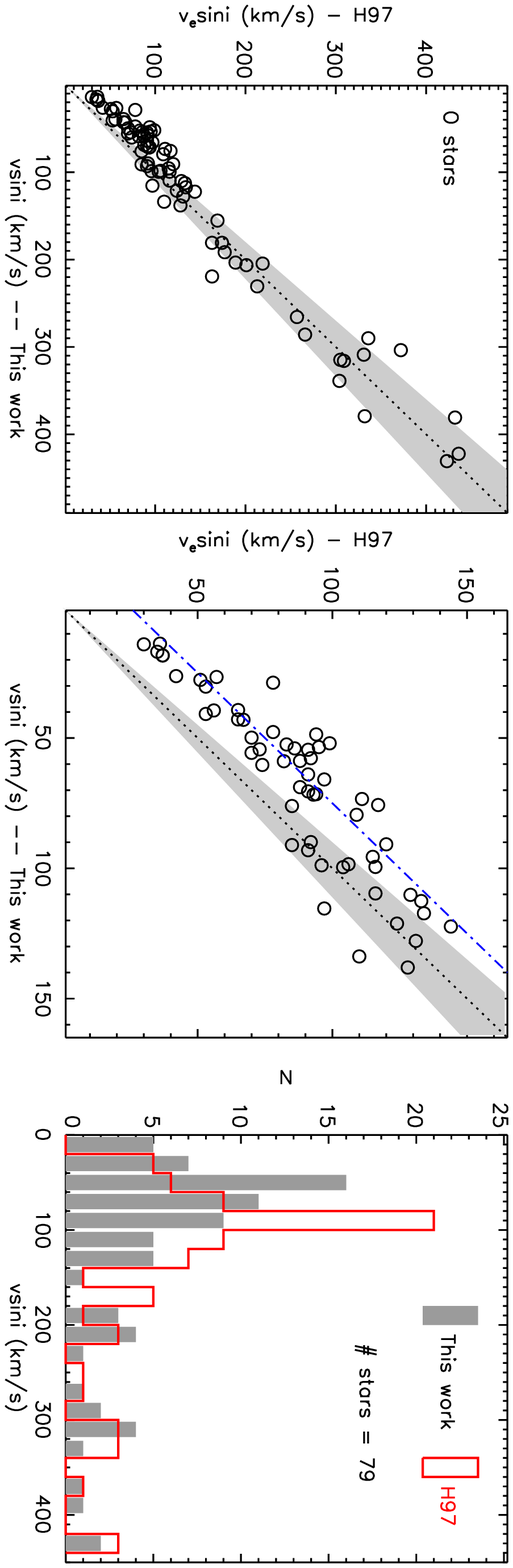}
\caption{Comparison with H97 (O stars in common). See Fig.~\ref{fig11} for 
explanations of the various panels.
}
\label{fig12}
\end{figure*}

\begin{figure*}[t!]
\centering
\includegraphics[width=5.8 cm,angle=90]{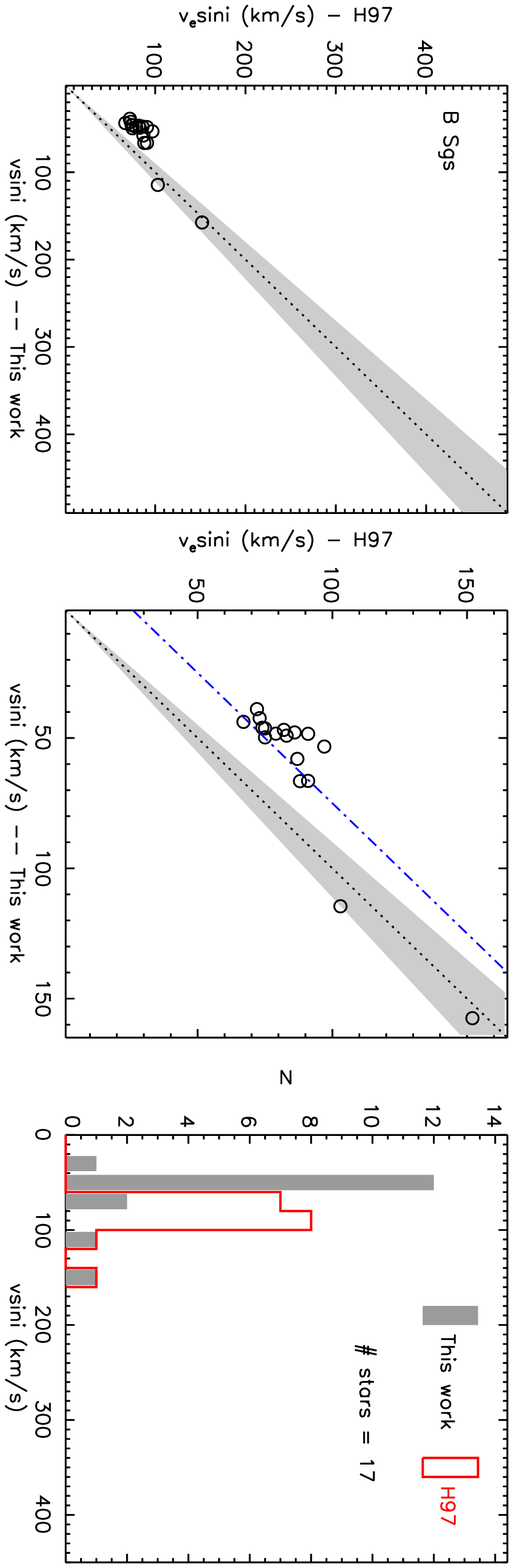}
\caption{Comparison with H97 (B~Sgs in common). See Fig.~\ref{fig11} for 
explanations of the various panels.
}
\label{fig13}
\end{figure*}

Comparison of results for the O stars and B\,Sgs samples is presented separately in
Figs. \ref{fig12} and \ref{fig13}, respectively. Once more, the effect of the non-rotational 
broadening on \vsini\ determinations is shown to be important not only in B\,Sgs, but 
also in O stars of all luminosity classes. Even O dwarfs with low \vsini\ are affected.

For the sake of completeness, we must remark here that the statement about the downward
revision of previous \vsini\ determinations is not applicable for the case of early-B dwarfs. 
As shown in Fig.~\ref{fig10}, the \macrot\ contribution to the line-profile is negligible 
compared to rotational broadening and hence a very good agreement between \vsini(FT) and
any measurement not accounting for \macrot\ broadening is expected.

\begin{figure*}[t!]
\centering
\includegraphics[width=12 cm,angle=90]{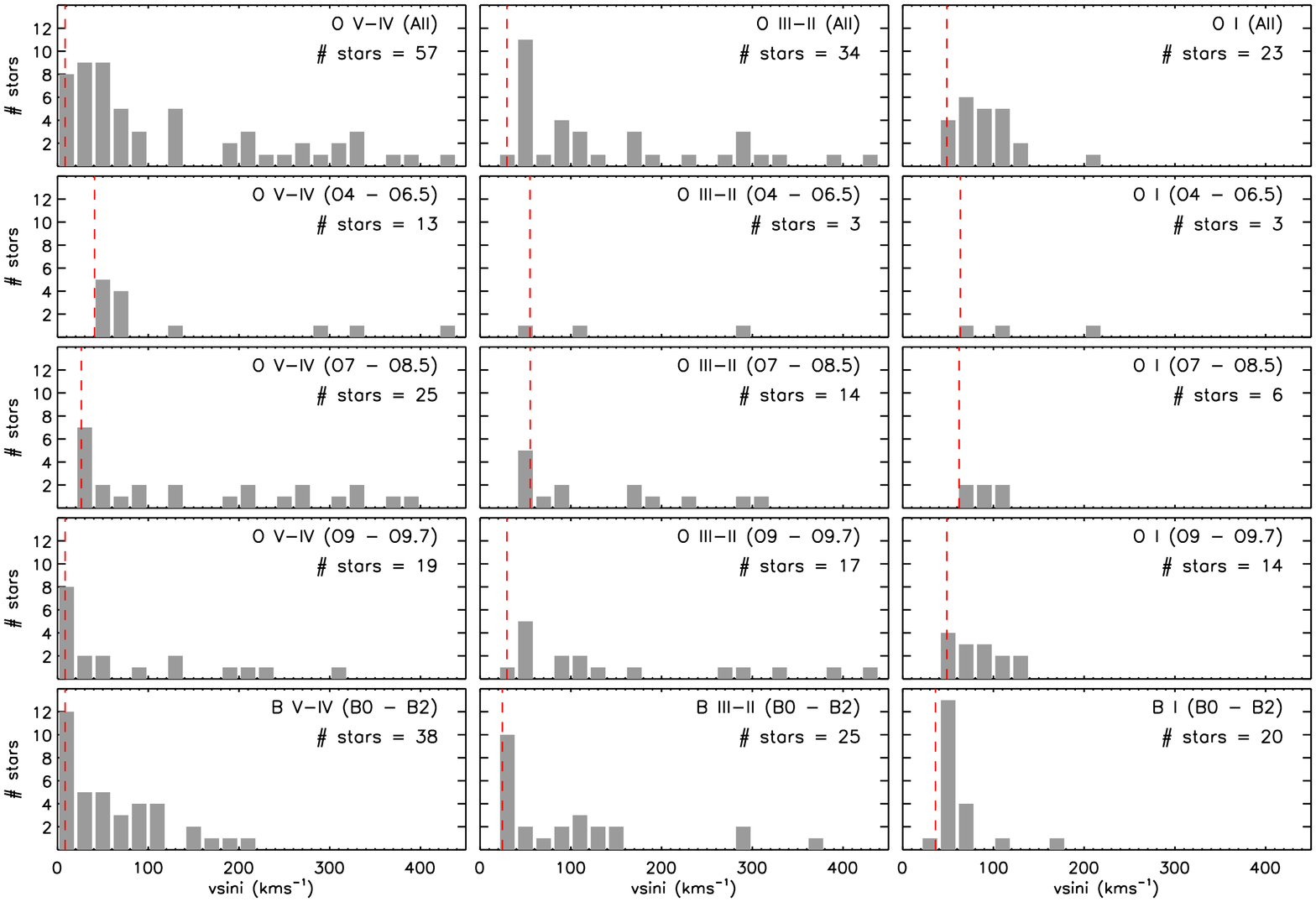}
\caption{Histograms resulting from the complete analyzed sample. From left to right:
dwarfs, (bright) giants, and supergiants. From top to bottom: all O stars, early-O, 
mid-O, late-O, and all early-B stars with a given luminosity class. Vertical dashed 
lines indicates the minimum \vsini\ measured in the specific sample (see also 
Fig \ref{fig15}).}
\label{fig14}
\end{figure*}

\subsubsection{The new \vsini\ distributions}\label{Sect432}

Fig.~\ref{fig14} groups the histograms resulting from the analysis of 
the global IACOB sample separated by spectral type (rows) and luminosity class (columns).
The row at the top includes all the O-type stars, while the bottom
row corresponds to the early B-type stars. Panels in between show the
distributions for the O stars separated in early, mid and late spectral 
types, respectively.

Before we start commenting on this figure it should be noted that the 
histogram corresponding to the early-B dwarfs has been only included for 
completeness, but must be left out from any discussion about \vsini\ distributions
since at present the 
IACOB spectroscopic database is totally biased towards (relatively) narrow 
line stars for these specific stars\footnote{The early-B dwarfs in the IACOB
spectroscopic database mainly come from the study of abundances in B-type
stars in Orion OB1 \citep{Sim+10, Nie11}
and the construction of an atlas of standard stars for spectral classification
at intermediate and high resolution.}. 
For a more complete overview of the distribution of projected rotational 
velocities in B-type stars in the Milky Way we refer to, e.g., \cite{Abt02}, 
\cite{Str05}, \cite{Hua06,Hua08}, \cite{Wol07}, \cite{Hua10}, \cite{Bra12}. 
In all these works it is shown that the distribution of projected rotational 
velocities of early-B dwarfs extends further to \vsini\ values higher than 200~\kms.

The number of stars in some of the histograms is not large enough to extract 
conclusions with the desired statistical significance; 
however, we can already highlight some trends (many of them in concurrence 
with previous studies by CE77, P96, and H97).
As already stated in previous works the global sample of Galactic O stars presents a 
bimodal distribution, containing a slow group with a \vsini\ peak near 40\,--\,60~\kms\, 
and a fast group extending up to $\sim$\,400~\kms. Interestingly, the low \vsini\
peak has been now revised downwards with respect to the 80\,--\,100~\kms\ indicated by
CE77, P96, and H97. This is basically an effect of the separation of the \macrot\
contribution to the total broadening (see Sect. \ref{Sect431}).

This bimodal distribution has been also recently obtained by \cite{Ram13}
in their study of rotational velocities in the O-star population of the 30\,Doradus
region. A similar analysis of 216 apparently single O-type stars in this star forming 
region of the Large Magellanic Cloud led to a low \vsini\ peak located at 40\,--\,80~\kms, 
and the high \vsini\ tail extending up to $\sim$\,600~\kms. In the case of stars in 30~Dor, 
the high \vsini\ tail extends further towards higher rotation speeds as expected 
from the lower metal content of the region (Z$_{30Dor}~\sim$\,0.5~Z$_{\odot}$). Unfortunately,
the accuracy of \vsini\ measurements in the study by \citeauthor{Ram13}
does not allow to firmly conclude if there is any difference in the location of the peak 
distribution.

\begin{table*}[t!]
\caption{Number of stars for various \vsini\ ranges of interest (see text).
The ranges in SpT correspond to those indicated in Figs.~\ref{fig14} and \ref{fig15}.}
\label{highvsini_table}
 \centering
\begin{tabular}{c|ccccc}
\hline
SpT \& LC & \multicolumn{4}{c}{\vsini\ [\kms]} \\
\hline
All LC     & $<$~50 & $<$~200 &  $>$~200 & $>$~300 & All \\
\hline
   O stars &  31 ( 26\%) & 90 ( 77\%) & 26 ( 22\%) & 12 ( 10\%) & 116  \\
Early B~Sgs & 10 ( 50\%) & 20 (100\%) &  0 (  0\%) &  0 (  0\%) &  20  \\
\noalign{\smallskip}
\hline
(V\,--\,IV)  & $<$~50 & $<$~200 &  $>$~200 & $>$~300 & All \\
\hline
     All O &  24 ( 42\%) & 41 ( 71\%) & 16 ( 28\%) &  8 ( 14\%) &  57  \\
    Early O &   3    & 10    &  3    &  2    & 13   \\
     Mid O &    9    & 15    & 10    &  5    & 25   \\
    Late O  &  12    & 16    &  3    &  1    & 19   \\
\noalign{\smallskip}
\hline
(III\,--\,II)  & $<$~50 & $<$~200 &  $>$~200 & $>$~300 & All \\
\hline
     All O &   4 ( 11\%) & 25 ( 73\%) &  9 ( 26\%) &  4 ( 11\%) &  34  \\
    Early O &   0    &  2    &  1    &  0    &  3   \\
      Mid O &   0    & 11    &  3    &  1    & 14   \\
    Late O &    4    & 12    &  5    &  3    & 17   \\
\noalign{\smallskip}
\hline
(I)        & $<$~50 & $<$~200 &  $>$~200 & $>$~300 & All \\
\hline
     All O &   1 (  4\%) & 22 ( 95\%) &  1 (  4\%) &  0 (  0\%) &  23  \\
    Early O &   0    &  2    &  1    &  0    &  3   \\
    Mid O  &    0    &  6    &  0    &  0    &  6   \\
    Late O &    1    & 14    &  0    &  0    & 14   \\
\hline
\end{tabular}
\end{table*}

CE77 indicated by that time that {\em we were somewhat at a loss to understand the 
apparent existence of a bimodal distribution for the main sequence O stars}. Nowadays, 
we count on a interesting scenario to explain this result \citep{dMi13}. In 
brief, these authors propose that the observed distribution may be a consequence of the 
combined effect of stellar winds, expansion, tides, mass trasfer, and mergers. In particular,
the high \vsini\ tail would mainly contain products resulting from massive star
binary interaction. For the purpose of comparison with the predictions by \citeauthor{dMi13} (or
any new proposed scenario), we indicate in Table \ref{highvsini_table} the number and
percentage of stars in various \vsini\ ranges of interest. It is specially
remarkable the good agreement between the predictions by \citeauthor{dMi13}, under the assumption
of constant star formation rate, and our percentages of observed O stars with \vsini\,$>$\,200 
and 300 \kms\ (23\% and 11\%, respectively). We should also note that these percentages are 
almost equal to those resulting from the studies by CE77, P96 and H97. This is a 
consequence of the fact that above $\sim$120~\kms\ the effect of the \macrot\ contribution 
to the global broadening is practically negligible when compared to rotational broadening 
(see Sect. \ref{Sect431}).

If we now split the distribution of projected rotational velocities in spectral type
and luminosity class it can be concluded that the
high velocity tail completely disappear in the case of O Sgs (see also Table
\ref{highvsini_table}), where stars concentrate
at \vsini\,=\,40\,--\,140~\kms. Interestingly, the distribution of B~Sgs (which are
considered the evolved stages of the mid and early O-type stars) also present a single peak
distribution concentrated in the low \vsini\ regime. This result was already pointed
out by H97; the only difference found here is that while H97 found the peaks of the two
distributions at $\sim$90 and 70~\kms\ (O and B~Sgs, respectively), our results led 
to 70~\kms, and 50~\kms, respectively. 

Another result that has attracted our attention refers to the comparison of \vsini\ distributions
corresponding to the evolutionary sequence starting in the mid-O~dwarfs and ending in the early-B~Sgs
(i.e. the diagonal going from the O7\,--\,O8.5~V\,--\,IV panel to the B0\,--\,B2~I panel). 
Independently of the origin of the high \vsini\ tail, the (almost) absence of early-B\,Sgs with 
projected rotational velocities larger than 100\,\kms, can be used as a observational evidence
that {\em evolution makes fast rotators to spin down very efficiently}. However, 
if the same mechanisms which make these stars to slow down affect all mid-O dwarfs, we would 
expect a larger number of early-B\,Sgs with lower \vsini. In fact, the peak of the distribution
of early-B~Sgs is located at somewhat higher \vsini\ values than that corresponding to the 
mid-O dwarfs. This result, also indicated by \cite{How04} in a more general context,
was used by previous authors as an indirect proof of the presence of some non-rotational 
macroscopic line broadening operating in massive stars. However, the techniques applied here
are supposed to provide actual projected rotational velocities. Therefore, either there is
something that we still do not understand from an evolutionary point of view (e.g. is the mechanism
reducing the surface rotation of massive stars different for the case of fast rotators and 
intermediate and low \vsini\ stars?), or our efforts to separate rotation from other sources 
of line-broadening are still failing in some cases.

In next section we discuss in more detail this (and other) remaining issues regarding the 
low \vsini\ region of the distributions, also providing a possible explanation and some guidelines
for future investigation. 


\subsubsection{The low \vsini\ regime}\label{Sect433}

\begin{figure}[t!]
\centering
\includegraphics[width=8. cm,angle=0]{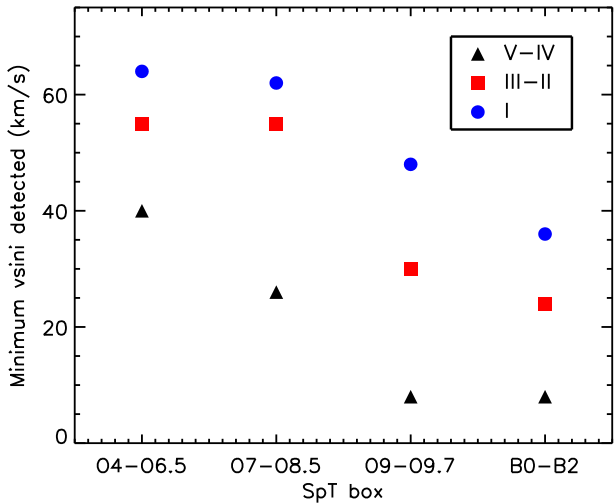}
\caption{Minimum \vsini\ measured for specific sub-samples, separated by SpT
and LC. The indicated values correspond to the dashed vertical lines in 
Fig.~\ref{fig14}. 
}
\label{fig15}
\end{figure}

The longstanding problem of the small number of low \vsini\ O-type stars and early-B~Sgs 
(CE77, P96, H97) seems now to be partially 
solved in view of the diagrams presented in Figs \ref{fig11}, \ref{fig12},
and \ref{fig13}. In particular, in Sect. \ref{Sect432}, we pointed out when comparing
with previous results by H97 that the number of O stars with measured \vsini\ values
below 40~\kms\ is now multiplied by a factor 2. However, looking closer to the 
\vsini\ distributions separated by spectral type and luminosity class 
(Fig. \ref{fig14}) one could conclude that, despite this improvement, there
still seems to be some limitation in the detectability of the very low \vsini\ stars. 
Apart from the evolutionary considerations indicated at the end of previous section, 
looking at Fig. \ref{fig15}, where we show the minimum detected \vsini\ in different 
spectral type boxes, and taking into account statistical arguments based on the probability to find
stars with a given inclination angle, it is remarkable that:

\begin{itemize}
\item No O\,Sgs with \vsini\,$<$\,50 \kms are still found\footnote{Actually, there is one 
star (HD\,37742), but is has \vsini\,=\,48~\kms.}. The situation is even more critical
for the case of mid and early O\,Sgs, where the lower \vsini\ measured is $\sim$\,60 \kms.
We should note, however, the very low statistics in this later case (only 7 stars analyzed).
\item While the relative number of B0\,--\,B2\,Sgs with measured projected rotational 
velocities below 50\,\kms\ is considerably larger ($\sim$\,50\%), the FT+GOF analysis
does not result in a \vsini$<$35 \kms\ for any of them, and only 2 from a total of 20 have
a \vsini$<$40~\kms. 
\item There is no problem with the relative number of low \vsini\ stars found in the case
of early-B and late-O dwarfs. There is a fair number of O9\,--\,B2 dwarfs in the IACOB
sample with measured \vsini\ below 40\,\kms, and the lower \vsini\ measured for this
category is of the order of 10 \kms.   
\item But for mid- and early-O dwarfs there seems to still be a limit in the minimum 
detected \vsini. This limit increases from $\sim$\,25~\kms\ in the case of mid-O dwarfs 
to $\sim$\,40~\kms\ for the dwarfs with spectral types earlier than O6.5. 
\item A similar situation occurs for the O stars with luminosity classes III and II
(this time also including the late O spectral types). Again, the lower limit in the
measured \vsini\ increases from $\sim$\,30~\kms to $\sim$\,55\,kms from the latest
to the earliest spectral types.
\item Finally, a low \vsini\ limit of $\sim$\,25 \kms\ is also found in the early-B 
(bright) giants
\end{itemize}

One possible explanation of this dependence of the minimum detected \vsini\ on spectral type
and luminosity class is related to the effect of microturbulence
on \vsini\ measurements. As indicated in Sect.~\ref{Sect34}, microturbulence can impose
a lower limit of detectability on the actual projected rotational velocity when
a strategy similar to the one considered in our study is applied. The rough estimates obtained
from the exercise presented in Sect.~\ref{Sect34} indicate that for a star with a
microturbulence of 20 \kms, any \vsini\,$\le$\,35 \kms, should be actually considered
as an upper limit. Since microturbulent velocities $\sim$\,15\,--\,20 \kms are typically 
found for early-B\,Sgs, this may explain why no B\,Sgs with \vsini\,$\le$\,40 \kms\ are 
still found. On the other hand, microturbulent velocities in the range 1\,--\,7~\kms\ and 
10\,--\,15 \kms\ (rough numbers!) are derived in the case of B dwarfs and giants, respectively. 
This would make the effect of microturbulence not so critical for B dwarfs 
(as indicated in Sect.~\ref{Sect34} if microturbulence is 5 \kms\ the upper \vsini\ limit moves down 
to $\sim$\,10 \kms), and intermediate for B giants.

What about mid and early O-type stars? Unfortunately, a direct estimation of microturbulent velocities from 
the analysis of observed spectra of mid and early O-type stars is not so straightforward 
as in the case of B-type and late O-type stars
Extrapolating the trends of microturbulence observed for B-type stars (where it is found that 
microturbulent velocities are a factor 2\,--\,3 larger in early-B than in late-B~Sgs) 
to the O-type stars domain could explain the tendency observed in Fig.~\ref{fig15}. However, this 
is by far too risky and, in addition, the associated microturbulent velocities expected for the 
earliest spectral types would be too large.

One could also consider the predictions by \cite{Can09}. These authors proposed 
that a physical connection may exist  between microturbulence in hot star atmospheres and 
sub-photospheric convective motions associated to the iron convection zone. Under this
hypothesis, they predict the regions in the HR diagram where larger values of 
microturbulent velocities should be expected. Interestingly, while the predictions for the 
late-O/early-B star domain roughly follow the observed tendency, values of microturbulence predicted 
for early and mid O-dwarfs are not expected to be specially large compared to later spectral 
types.

\section{Summary and future prospects}\label{Sect5}

In this paper we reassess previous determinations of projected rotational 
velocities (\vsini) in Galactic OB stars using a large, high quality spectroscopic
dataset (drawn from the {\em IACOB spectroscopic database of Northern Galactic OB stars}) 
and a powerful technique which accounts for other sources of broadening appart from
rotation affecting the diagnostic lines. In particular, we investigate the effect of 
\macrot\ and microturbulent broadenings on \vsini\ measurements.

Motivated by the investigation presented in this paper we have developed a
versatile and user friendly IDL tool --- based on a combined Fourier
transform (FT) + goodness of fit (GOF) methodology --- for the line-broadening 
characterization in OB-type stars: the {\tt iacob-broad}. The procedure allows
it to extract information about \vsini\ and the \macrot\ broadening 
(\vmacro) from the stellar line-profiles under a variety of situations.
The {\tt iacob-broad} tool has been used for three purposes:

\begin{enumerate}
 \item The investigation of the effect that the assumed \macrot\ profile (either
 isotropic Gaussian or radial-tangential) has on the GOF-based \vsini\ 
 and \vmacro\ estimations.
 \item The investigate of the possible consequences of neglecting the effect 
 of microturbulence on the determination of these two quantities
 \item The determination of \vsini\ and the size of the \macrot\ broadening in 
 a sample of $\sim$\,200 Galactic OB-type stars with spectral types ranging from 
 O4 to B2 and covering all luminosity classes. In particular, we wanted to
 compare the derived \vsini\ with previous determinations not accounting for
 the \macrot\ broadening contribution.
 \end{enumerate}

We show that \vsini\ and \vmacro\ determined by means of the GOF method critically
depend on the assumed \macrot\ profile, and present further observational 
evidence supporting a previous statement quoted in \cite{Sim10} indicating that 
a radial-tangential description is better suited to infer GOF-based projected 
rotational velocities. While the derived \vsini(GOF,\TRT) values are in perfect 
agreement with those resulting from the FT analysis, the isotropic Gaussian 
profile systematically led to much lower values of \vsini.

The very good agreement found between \vsini(FT) and \vsini(GOF, \TRT) from
the analysis of our high resolution, high S/N spectroscopic dataset suggests
that for worse quality datasets (e.g. fainter Galactic OB stars, or extragalactic
objects) we can adopt the values provided by the GOF method, as it is less sensitive
to noise than the FT method, as far as a radial-tangential formulation of the
\macrot\ profile is adopted.

We also warn about the danger of combining measurements of the \macrot\ 
velocity provided by different authors in the literature if they are obtained
by assuming a different \macrot\ profile. 

Following the guidelines by \cite{Gra76}, we show that any \vsini\ and \TRT\
measurement in OB-type stars (either from FT of GOF) below $\sim$\,40 \kms\ 
must be considered as an upper limit if the effects of microturbulence are not 
taken into account in the line-broadening analysis. 

An important outcome from our study is the confirmation that the \macrot\ broadening
is present not only in B~Sgs, but also in O-type stars of all luminosity classes
(as the analysis of smaller samples had already began to show). As a consequence,
those previous determinations of \vsini\ in these type of stars asumming that rotation is the sole source
of line-broadening below $\sim$\,120~\kms need to be systematically revised 
downwards\footnote{This statement is not applicable to early-B dwarfs, where
the \macrot\ broadening contribution to the line profiles is negligible.}
by 25 ($\pm$20)~\kms. This implies important modifications of the distributions
of projected rotational velocities in Galactic OB-type stars. In particular the
low \vsini\ peak previously found in 80\,--\,100~\kms\ (in the global \vsini\ 
distributions) has now been revised downwards to 40\,--\,60~\kms.

Although the longstanding problem of the small number of low \vsini\ O-type stars
and early-B~Sgs has been now partially solved, there seems to still remain some
limitation in the detectability of the very low \vsini\ stars, specially for
the case of mid and early O stars and the early-B~Sgs. We have found some interesting
trends of the minimum detected \vsini\ in our analized sample of Galactic stars
with spectral type and luminosity class. We indicate that this result
could be an effect due to the limitations of the FT+GOF strategy when
microturbulence is not taken into account.

It is still too premature to conclude that the effect of microturbulence on
\vsini\ measurements is the cause of this observational result; however, this is an interesting
possibility that should be explored in more detail in the future. As a guide we propose 
three different aspects to be investigated: (a) from a methodological point of view, 
is it possible to disentangle the spectroscopic features (either in the wavelength or 
the Fourier domains) associated to microturbulence from those exclusively produced 
by rotation?, (b) from an observational point of view, can we increase our knowledge 
about the microturbulence in O-type stars?, and (c) from a theoretical/modeling point 
of view, explore further the scenario proposed by \cite{Can09} and compare the
corresponding predictions with observational constraints. 

In addition, the extension of this type of study to other metallicities (also increasing
the number of analyzed stars in the Galaxy with the incorporation of the Southern
sample), and the exploration of other possible effects which could invalidate our 
state-of-the-art techniques used to measure projected rotational velocities in O 
and B-type stars (under certain circumstances) are also warranted.


%
\begin{acknowledgements}

This work has been funded by the Spanish Ministry of Economy and Competitiveness 
(MINECO) under the grants AYA2010-21697-C05-04, Consolider-Ingenio 2010 CSD2006-00070,
and Severo Ochoa SEV-2011-0187, and by the Canary Islands Government under grant 
PID2010119. SS-D kindly acknowledge the staff at the Nordic Optical Telescope for
their professional competence and always useful help during more than 50 observing 
nights between 2008 and 2013. Also to my observing colleagues I. Negueruela,
J. Lorenzo, N. Castro and M. Garcia. We are extremely grateful to N. Langer and the 
referee, F. Royer, for the time devoted to read the first version of the paper 
and his very useful and constructive comments.

\end{acknowledgements}


%

\appendix

\section{Tables}

\begin{table*}[t!]
\caption{Results from the {\tt iacob-broad} analysis of the O and B Supergiants (LC I).
}\label{taba1}
\centering
\begin{tabular}{lrllccccc}
\hline \hline
\noalign{\smallskip}
HD number & SpT & LC & Line & EW     & S/N & \vsini (FT) &  \vsini (GOF) & \vmacro (GOF) \\
          &     &    &      & [m\AA] &     & [\kms] & [\kms] & [\kms]  \\
\hline
\noalign{\smallskip}
    ...    &       ...  &        ... &       ...  &  ... & ... & ... & ... & ... \\
   HD91316 &         B1 &    IabNstr &      SiIII & 425 & 370 &  49 &  49 &  72  \\
    ...    &       ...  &        ... &       ...  &  ... & ... & ... & ... & ... \\
\noalign{\smallskip}
\hline \hline
\end{tabular}
\end{table*}
\begin{table*}[t!]
\caption{Results from the {\tt iacob-broad} analysis of the O and B (bright) giants (LCs II and III).
}\label{taba3}
\centering
\begin{tabular}{lrllccccc}
\hline \hline
\noalign{\smallskip}
HD number & SpT & LC & Line & EW     & S/N & \vsini (FT) &  \vsini (GOF) & \vmacro (GOF) \\
          &     &    &      & [m\AA] &     & [\kms] & [\kms] & [\kms]  \\
\hline
\noalign{\smallskip}
    ...    &       ...  &        ... &       ...  &  ... & ... & ... & ... & ... \\
   HD36861 &         O8 &   III((f)) &       OIII & 301 & 292 &  57 &  60 &  60  \\
    ...    &       ...  &        ... &       ...  &  ... & ... & ... & ... & ... \\
\noalign{\smallskip}
\hline \hline
\end{tabular}
\end{table*}
\begin{table*}[t!]
\caption{Results from the {\tt iacob-broad} analysis of the O dwarfs and subgiants (LC V and IV).
}\label{taba4}
\centering
\begin{tabular}{lrllccccc}
\hline \hline
\noalign{\smallskip}
HD number & SpT & LC & Line & EW     & S/N & \vsini (FT) &  \vsini (GOF) & \vmacro (GOF)\\
          &     &    &      & [m\AA] &     & [\kms] & [\kms] & [\kms] \\
\hline
\noalign{\smallskip}
    ...    &       ...  &        ... &       ...  &  ... & ... & ... & ... & ... \\
   HD46966 &       O8.5 &         IV &       OIII & 172 & 244 &  39 &  39 &  68  \\
    ...    &       ...  &        ... &       ...  &  ... & ... & ... & ... & ... \\
\noalign{\smallskip}
\hline \hline
\end{tabular}
\end{table*}
\begin{table*}[t!]
\caption{Results from the {\tt iacob-broad} analysis of the B dwarfs and subgiants (LC V and IV).
}\label{taba5}
\centering
\begin{tabular}{lrllccccc}
\hline \hline
\noalign{\smallskip}
HD number & SpT & LC & Line & EW     & S/N & \vsini (FT) &  \vsini (GOF) & \vmacro (GOF) \\
          &     &    &      & [m\AA] &     & [\kms] & [\kms] & [\kms] \\
\hline
\noalign{\smallskip}
    ...    &       ...  &        ... &       ...  &  ... & ... & ... & ... & ... \\
   HD37042 &       B0.7 &          V &      SiIII & 149 & 217 &  33 &  33 &  13  \\
    ...    &       ...  &        ... &       ...  &  ... & ... & ... & ... & ... \\
\noalign{\smallskip}
\hline \hline
\end{tabular}
\end{table*}
\begin{table*}[t!]
\caption{Results from the {\tt iacob-broad} analysis of the O and B stars identified as SB1. The two
O\,f?p stars HD\,108 and HD\,191612 are also included in this table.
}\label{taba6}
\centering
\begin{tabular}{lrllccccc}
\hline \hline
\noalign{\smallskip}
HD number & SpT & LC & Line & EW     & S/N & \vsini (FT) &  \vsini (GOF) & \vmacro (GOF) \\
          &     &    &      & [m\AA] &     & [\kms] & [\kms] & [\kms] \\
\hline
\noalign{\smallskip}
    ...    &       ...  &        ... &       ...  &  ... & ... & ... & ... & ... \\
  HD199579 &       O6.5 &    V((f))z &       OIII & 211 & 215 &  49 &  54 &  79  \\
    ...    &       ...  &        ... &       ...  &  ... & ... & ... & ... & ... \\
\noalign{\smallskip}
\hline \hline
\end{tabular}
\end{table*}

%
%
\end{document}